\documentclass{conm-p-l}
\usepackage{amsmath,amssymb,amsthm}
\usepackage{pstricks}
\usepackage{graphicx}

\newcommand{\propagator}{D}
\newcommand{\deri}{A}
\newcommand{\opro}{\circ}
\newcommand{\Gzero}{G^0}

\newcommand{\rhoc}{\rho_c}
\newcommand{\rhoi}{\rho_{I}}
\newcommand{\Hint}{H_{\mathrm{int}}}
\newcommand{\PsiGL}{\Psi_{\mathrm{GL}}}

\newcommand{\SYM}{\mathbb Sym}

\newcommand{\ee}{{\mathrm{e}}}
\newcommand{\dd}{{\mathrm{d}}}

\newcommand{\bfr}{{\mathbf{r}}}

\newcommand{\mathcalH}{{\mathcal H}}

\newcommand{\R}{{\mathbb{R}}}

\newcommand{\counit}{{\varepsilon}}

\newcommand{\ix}[1]{{}_{\scriptscriptstyle(#1)}}

\newcommand{\Deltas}{\delta}
\newcommand{\Deltad}{{\delta}}
\newcommand{\Deltadu}{{\underline{\Deltad}}}

\newcommand{\Deltau}{{\underline{\Delta}}}

\newcommand{\is}[1]{{}_{\scriptscriptstyle\{ #1 \}}}
\newcommand{\isu}[1]{{}_{\scriptscriptstyle\{\underline #1 \}}}
\newcommand{\iis}[1]{{}_{{\scriptscriptstyle{\{}}{\scriptstyle{#1}}
          {\scriptscriptstyle{\}}} }}

\newcommand{\ii}[1]{{}_{{\scriptscriptstyle{(}}{\scriptstyle{#1}}
          {\scriptscriptstyle{)}} }}
\newcommand{\iu}[1]{{}_{\scriptscriptstyle(\underline{#1})}}
\newcommand{\iiu}[1]{{}_{\scriptstyle(\underline{#1})}}
\newcommand{\id}{\mathrm{id}}

\newcommand{\phiS}{\varphi_S}

\newcommand{\ket}[1]{| #1 \rangle}
\newcommand{\bra}[1]{\langle #1 |}

\newcommand{\C}{\mathbb{C}}
\newcommand{\CC}{\mathbb{C}}
\newcommand{\N}{\mathbb{N}}
\newcommand{\NN}{\mathbb{N}}

\newtheorem{prop}{Proposition}[section]
\newtheorem{lem}[prop]{Lemma}

\newtheorem{thm}{Theorem}[section]
\newtheorem{de}[prop]{Definition}

\newcommand{\tunn}{\setlength{\unitlength}{3pt}
\psset{unit=3pt}
\psset{runit=3pt}
\psset{linewidth=0.2}
\begin{pspicture}(0,0)(2,2)
\psdots[dotstyle=*](1,1)
\end{pspicture}}

\newcommand{\tdeux}{\setlength{\unitlength}{3pt}
\psset{unit=3pt}
\psset{runit=3pt}
\psset{linewidth=0.2}
\begin{pspicture}(0,0)(3,2)
\psline(0,1)(3,1)
\psdots[dotstyle=*](0,1)
\psdots[dotstyle=o](3,1)
\end{pspicture}}

\newcommand{\ttroisn}{\setlength{\unitlength}{3pt}
\psset{unit=3pt}
\psset{runit=3pt}
\psset{linewidth=0.2}
\begin{pspicture}(0,0)(5,2)
\psline(0,1)(2.5,1)
\psline(2.5,1)(5,1)
\psdots[dotstyle=*](0,1)
\psdots[dotstyle=o](2.5,1)
\psdots[dotstyle=*](5,1)
\end{pspicture}}

\newcommand{\ttroisb}{\setlength{\unitlength}{3pt}
\psset{unit=3pt}
\psset{runit=3pt}
\psset{linewidth=0.2}
\begin{pspicture}(0,0)(5,2)
\psline(0,1)(2.5,1)
\psline(2.5,1)(5,1)
\psdots[dotstyle=o](0,1)
\psdots[dotstyle=*](2.5,1)
\psdots[dotstyle=o](5,1)
\end{pspicture}}

\newcommand{\tquatren}{\setlength{\unitlength}{3pt}
\psset{unit=3pt}
\psset{runit=3pt}
\psset{linewidth=0.2}
\begin{pspicture}(0,0)(4,3)
\psline(0,0)(2,1)
\psline(4,0)(2,1)
\psline(2,3)(2,1)
\psdots[dotstyle=*](0,0)
\psdots[dotstyle=*](4,0)
\psdots[dotstyle=*](2,3)
\psdots[dotstyle=o](2,1)
\end{pspicture}}

\newcommand{\tquatrel}{\setlength{\unitlength}{3pt}
\psset{unit=3pt}
\psset{runit=3pt}
\psset{linewidth=0.2}
\begin{pspicture}(0,0)(7.5,2)
\psline(0,1)(2.5,1)
\psline(2.5,1)(5,1)
\psline(5,1)(7.5,1)
\psdots[dotstyle=*](0,1)
\psdots[dotstyle=o](2.5,1)
\psdots[dotstyle=*](5,1)
\psdots[dotstyle=o](7.5,1)
\end{pspicture}}

\newcommand{\tquatreb}{\setlength{\unitlength}{3pt}
\psset{unit=3pt}
\psset{runit=3pt}
\psset{linewidth=0.2}
\begin{pspicture}(0,0)(4,3)
\psline(0,0)(2,1)
\psline(4,0)(2,1)
\psline(2,3)(2,1)
\psdots[dotstyle=o](0,0)
\psdots[dotstyle=o](4,0)
\psdots[dotstyle=o](2,3)
\psdots[dotstyle=*](2,1)
\end{pspicture}}

\newcommand{\tcinqn}{\setlength{\unitlength}{3pt}
\psset{unit=3pt}
\psset{runit=3pt}
\psset{linewidth=0.2}
\begin{pspicture}(0,0)(4,3)
\psline(0,0)(2,1.5)
\psline(4,0)(2,1.5)
\psline(0,3)(2,1.5)
\psline(4,3)(2,1.5)
\psdots[dotstyle=*](0,0)
\psdots[dotstyle=*](4,0)
\psdots[dotstyle=*](0,3)
\psdots[dotstyle=*](4,3)
\psdots[dotstyle=o](2,1.5)
\end{pspicture}}

\newcommand{\tcinqln}{\setlength{\unitlength}{3pt}
\psset{unit=3pt}
\psset{runit=3pt}
\psset{linewidth=0.2}
\begin{pspicture}(0,0)(10,2)
\psline(0,1)(2.5,1)
\psline(2.5,1)(5,1)
\psline(5,1)(7.5,1)
\psline(7.5,1)(10,1)
\psdots[dotstyle=*](0,1)
\psdots[dotstyle=o](2.5,1)
\psdots[dotstyle=*](5,1)
\psdots[dotstyle=o](7.5,1)
\psdots[dotstyle=*](10,1)
\end{pspicture}}

\newcommand{\tcinqfn}{\setlength{\unitlength}{3pt}
\psset{unit=3pt}
\psset{runit=3pt}
\psset{linewidth=0.2}
\begin{pspicture}(0,0)(7.5,2)
\psline(0,0)(2.5,1)
\psline(0,2)(2.5,1)
\psline(2.5,1)(5,1)
\psline(5,1)(7.5,1)
\psdots[dotstyle=*](0,0)
\psdots[dotstyle=*](0,2)
\psdots[dotstyle=o](2.5,1)
\psdots[dotstyle=*](5,1)
\psdots[dotstyle=o](7.5,1)
\end{pspicture}}

\newcommand{\tcinqfb}{\setlength{\unitlength}{3pt}
\psset{unit=3pt}
\psset{runit=3pt}
\psset{linewidth=0.2}
\begin{pspicture}(0,0)(7.5,2)
\psline(0,0)(2.5,1)
\psline(0,2)(2.5,1)
\psline(2.5,1)(5,1)
\psline(5,1)(7.5,1)
\psdots[dotstyle=o](0,0)
\psdots[dotstyle=o](0,2)
\psdots[dotstyle=*](2.5,1)
\psdots[dotstyle=o](5,1)
\psdots[dotstyle=*](7.5,1)
\end{pspicture}}

\newcommand{\tcinqlb}{\setlength{\unitlength}{3pt}
\psset{unit=3pt}
\psset{runit=3pt}
\psset{linewidth=0.2}
\begin{pspicture}(0,0)(10,2)
\psline(0,1)(2.5,1)
\psline(2.5,1)(5,1)
\psline(5,1)(7.5,1)
\psline(7.5,1)(10,1)
\psdots[dotstyle=o](0,1)
\psdots[dotstyle=*](2.5,1)
\psdots[dotstyle=o](5,1)
\psdots[dotstyle=*](7.5,1)
\psdots[dotstyle=o](10,1)
\end{pspicture}}

\newcommand{\tcinqb}{\setlength{\unitlength}{3pt}
\psset{unit=3pt}
\psset{runit=3pt}
\psset{linewidth=0.2}
\begin{pspicture}(0,0)(4,3)
\psline(0,0)(2,1.5)
\psline(4,0)(2,1.5)
\psline(0,3)(2,1.5)
\psline(4,3)(2,1.5)
\psdots[dotstyle=o](0,0)
\psdots[dotstyle=o](4,0)
\psdots[dotstyle=o](0,3)
\psdots[dotstyle=o](4,3)
\psdots[dotstyle=*](2,1.5)
\end{pspicture}}

\newcommand{\tsixn}{\setlength{\unitlength}{3pt}
\psset{unit=3pt}
\psset{runit=3pt}
\psset{linewidth=0.2}
\begin{pspicture}(0,0)(3.8,3.6)
\psline(1.9,3.6)(1.9,1.6)
\psline(3.8,2.2)(1.9,1.6)
\psline(3.1,0)(1.9,1.6)
\psline(0.7,0)(1.9,1.6)
\psline(0,2.2)(1.9,1.6)
\psdots[dotstyle=*](1.9,3.6)
\psdots[dotstyle=*](3.8,2.2)
\psdots[dotstyle=*](3.1,0)
\psdots[dotstyle=*](0.7,0)
\psdots[dotstyle=*](0,2.2)
\psdots[dotstyle=o](1.9,1.6)
\end{pspicture}}

\newcommand{\tsixfn}{\setlength{\unitlength}{3pt}
\psset{unit=3pt}
\psset{runit=3pt}
\psset{linewidth=0.2}
\begin{pspicture}(0,0)(10,2)
 \psline(0,0)(2.5,1)
\psline(0,2)(2.5,1)
\psline(2.5,1)(5,1)
\psline(5,1)(7.5,1)
\psline(7.5,1)(10,1)
\psdots[dotstyle=*](0,0)
\psdots[dotstyle=*](0,2)
\psdots[dotstyle=o](2.5,1)
\psdots[dotstyle=*](5,1)
\psdots[dotstyle=o](7.5,1)
\psdots[dotstyle=*](10,1)
\end{pspicture}}

\newcommand{\tsixgfn}{\setlength{\unitlength}{3pt}
\psset{unit=3pt}
\psset{runit=3pt}
\psset{linewidth=0.2}
\begin{pspicture}(0,0)(7.5,2)
\psline(0,0)(2.5,0)
\psline(0,2)(2.5,2)
\psline(2.5,0)(5,1)
\psline(2.5,2)(5,1)
\psline(5,1)(7.5,1)
\psdots[dotstyle=*](0,0)
\psdots[dotstyle=*](0,2)
\psdots[dotstyle=o](2.5,0)
\psdots[dotstyle=o](2.5,2)
\psdots[dotstyle=*](5,1)
\psdots[dotstyle=o](7.5,1)
\end{pspicture}}

\newcommand{\tsixftn}{\setlength{\unitlength}{3pt}
\psset{unit=3pt}
\psset{runit=3pt}
\psset{linewidth=0.2}
\begin{pspicture}(0,0)(7.5,3)
\psline(0,0)(2.5,1.5)
\psline(0,1.5)(2.5,1.5)
\psline(0,3)(2.5,1.5)
\psline(2.5,1.5)(5,1.5)
\psline(5,1.5)(7.5,1.5)
\psdots[dotstyle=*](0,0)
\psdots[dotstyle=*](0,1.5)
\psdots[dotstyle=*](0,3)
\psdots[dotstyle=o](2.5,1.5)
\psdots[dotstyle=*](5,1.5)
\psdots[dotstyle=o](7.5,1.5)
\end{pspicture}}

\newcommand{\tsixff}{\setlength{\unitlength}{3pt}
\psset{unit=3pt}
\psset{runit=3pt}
\psset{linewidth=0.2}
\begin{pspicture}(0,0)(7.5,2)
\psline(0,0)(2.5,1)
\psline(0,2)(2.5,1)
\psline(2.5,1)(5,1)
\psline(5,1)(7.5,0)
\psline(5,1)(7.5,2)
\psdots[dotstyle=*](0,0)
\psdots[dotstyle=*](0,2)
\psdots[dotstyle=o](2.5,1)
\psdots[dotstyle=*](5,1)
\psdots[dotstyle=o](7.5,0)
\psdots[dotstyle=o](7.5,2)
\end{pspicture}}

\newcommand{\tsixl}{\setlength{\unitlength}{3pt}
\psset{unit=3pt}
\psset{linewidth=0.2}
\begin{pspicture}(0,0)(12.5,2)
\psline(0,1)(2.5,1)
\psline(2.5,1)(5,1)
\psline(5,1)(7.5,1)
\psline(7.5,1)(10,1)
\psline(10,1)(12.5,1)
\psdots[dotstyle=*](0,1)
\psdots[dotstyle=o](2.5,1)
\psdots[dotstyle=*](5,1)
\psdots[dotstyle=o](7.5,1)
\psdots[dotstyle=*](10,1)
\psdots[dotstyle=o](12.5,1)
\end{pspicture}}

\newcommand{\tsixftb}{\setlength{\unitlength}{3pt}
\psset{unit=3pt}
\psset{runit=3pt}
\psset{linewidth=0.2}
\begin{pspicture}(0,0)(7.5,3)
\psline(0,0)(2.5,1.5)
\psline(0,1.5)(2.5,1.5)
\psline(0,3)(2.5,1.5)
\psline(2.5,1.5)(5,1.5)
\psline(5,1.5)(7.5,1.5)
\psdots[dotstyle=o](0,0)
\psdots[dotstyle=o](0,1.5)
\psdots[dotstyle=o](0,3)
\psdots[dotstyle=*](2.5,1.5)
\psdots[dotstyle=o](5,1.5)
\psdots[dotstyle=*](7.5,1.5)
\end{pspicture}}

\newcommand{\tsixgfb}{\setlength{\unitlength}{3pt}
\psset{unit=3pt}
\psset{runit=3pt}
\psset{linewidth=0.2}
\begin{pspicture}(0,0)(7.5,2)
\psline(0,0)(2.5,0)
\psline(0,2)(2.5,2)
\psline(2.5,0)(5,1)
\psline(2.5,2)(5,1)
\psline(5,1)(7.5,1)
\psdots[dotstyle=o](0,0)
\psdots[dotstyle=o](0,2)
\psdots[dotstyle=*](2.5,0)
\psdots[dotstyle=*](2.5,2)
\psdots[dotstyle=o](5,1)
\psdots[dotstyle=*](7.5,1)
\end{pspicture}}

\newcommand{\tsixfb}{\setlength{\unitlength}{3pt}
\psset{unit=3pt}
\psset{runit=3pt}
\psset{linewidth=0.2}
\begin{pspicture}(0,0)(10,2)
\psline(0,0)(2.5,1)
\psline(0,2)(2.5,1)
\psline(2.5,1)(5,1)
\psline(5,1)(7.5,1)
\psline(7.5,1)(10,1)
\psdots[dotstyle=o](0,0)
\psdots[dotstyle=o](0,2)
\psdots[dotstyle=*](2.5,1)
\psdots[dotstyle=o](5,1)
\psdots[dotstyle=*](7.5,1)
\psdots[dotstyle=o](10,1)
\end{pspicture}}
\newcommand{\tsixb}{\setlength{\unitlength}{3pt}
\psset{unit=3pt}
\psset{runit=3pt}
\psset{linewidth=0.2}
\begin{pspicture}(0,0)(3.8,3.6)
\psline(1.9,3.6)(1.9,1.6)
\psline(3.8,2.2)(1.9,1.6)
\psline(3.1,0)(1.9,1.6)
\psline(0.7,0)(1.9,1.6)
\psline(0,2.2)(1.9,1.6)
\psdots[dotstyle=o](1.9,3.6)
\psdots[dotstyle=o](3.8,2.2)
\psdots[dotstyle=o](3.1,0)
\psdots[dotstyle=o](0.7,0)
\psdots[dotstyle=o](0,2.2)
\psdots[dotstyle=*](1.9,1.6)
\end{pspicture}}

\begin{document}
% 
%\begin{titlepage}
% 
\title{One-particle irreducibility with initial correlations}
\author{Christian Brouder}
\address{Institut de Min\'eralogie et de Physique des Milieux
  Condens\'es, CNRS UMR7590,
 Universit\'es Paris 6 et 7, IPGP, 140 rue de Lourmel,
  75015 Paris, France.}
\author{Fr\'ed\'eric Patras}
\address{Universit\'e de Nice,
Laboratoire J.-A. Dieudonn\'e, CNRS UMR 6621,
Parc Valrose, 06108 Nice Cedex 02, France.}

\begin{abstract}
In quantum field theory (QFT),
the vacuum expectation of a normal product of creation and
annihilation operators is always zero. This simple property paves the way to the classical treatment of perturbative QFT. This is no longer the case in the presence of initial correlations, that is if the
vacuum is replaced by a general state. As a consequence, the
combinatorics of correlated systems such as the ones occurring in many-body physics is more complex than that of
quantum field theory and the general theory has made very slow progress. 
Similar observations hold in statistical physics or quantum probability for the perturbation series arising from the study of non Gaussian measures. In
this work, an analysis of the Hopf algebraic aspects of quantum field
theory is used to derive the structure of Green functions in terms
of connected and one-particle irreducible Green functions for perturbative QFT in the presence of initial correlations.
\end{abstract}

\maketitle
%\end{titlepage}

\section{Introduction}

In quantum field theory (QFT), the initial state is most
often the vacuum. Many quantum field concepts,
such as Feynman diagrams, the Dyson equation
and the Bethe-Salpeter equation rest on the 
special properties of the vacuum. These desirable
concepts extend to special states called
\emph{quasi-free states} \cite{Robinson,Kay1,Kay2}.

For general initial states, it is not possible to write the
Green function in terms of standard Feynman diagrams
and the structure of the Green functions is more complex.
For example, the Dyson and Bethe-Salpeter equations
do not hold.
The Dyson equation describes the structure of the
two-point interacting Green function. Its extension to
non-quasi-free states was discovered by Hall~\cite{Hall} in 1975.
However, the equivalent structural equations for $2n$-point
interacting Green functions (with $n>1$) is not know for general 
initial states. The determination of this
structure is the main purpose of the present article, with a particular emphasis on the notion of one-particular-irreducibility in this framework.

We stress that the problem of the
calculation of Green functions for initial states
that are not quasi-free has important applications.
For instance, many highly-correlated 
materials contain transition metals where states of the
$3d$ shell are degenerate. The consequence of this degeneracy 
is that a small external 
perturbation can create a very strong change in the state of the
system. For instance, a small external magnetic field induces
a strong variation in resistance (giant magnetoresistance), that
is used to build high-density storage disks. The knowledge of the
Green functions would enable us to calculate accurately
the properties of such materials.

Although we will use mainly the language of QFT, let us mention that the 
situation where the vacuum is the initial state corresponds in statistical 
physics to the case of Euclidean measures as showing up e.g. in Euclidean 
Quantum Field Theory. This is because of the joint use of the Wick theorem 
(in its simplest, Gaussian, form). For general measures, the problem of 
determining the fine combinatorial structure of perturbation series was 
addressed in \cite{Djah} where, in particular, the combinatorics of truncated 
(or connected) moment functions was studied intensively. We refer to this 
article, also for further motivations and examples of computations of 
thermodynamic limits involving the use of generalized Feynman diagrams as 
we also use them. 

Concretely, in the present article, we investigate the structure of Green 
functions with Hopf algebraic methods. 
Hopf algebras have been implicitly used for a long time in
statistical physics and quantum field theory: 
Ruelle~\cite{Ruelle}, Borchers~\cite{Borchers} and 
Stora~\cite{Stora-73} used 
a product that is called the \emph{convolution product}
in the Hopf language. Wightman and Challifour \cite{Wightman-70}
defined a \emph{triple dot product} that was rediscovered
only much later in quantum chemistry~\cite{Kutzelnigg}
and was given an algebraic meaning in \cite{BrouderQG},
where it was called \emph{adapted normal product}.
The work of Wightman and Challifour was summarized by 
Stora in ref. \cite{Stora-73}.

The explicit introduction of Hopf algebras
by Kreimer and Connes at the level of trees and 
Feynman diagrams \cite{Kreimer98,Connes} sparked a
reformulation of many quantum field constructions
(renormalization \cite{CKI,CKII,EGP1}, Wick's theorem
\cite{Fauser}, quantization \cite{BrouderQG,Hirshfeld},
structure of Green functions \cite{Mestre,Mestre2},
gauge theory \cite{Suijlekom-07}).
The result of these efforts is a reasonably complete presentation of 
QFT in terms of Hopf algebraic concepts
\cite{BrouderMN}.
Hopf algebras, which are powerful tools to solve
combinatorial problems \cite{Joni}, could be expected to help also in 
the presence of a general state. 
Indeed, the use of Hopf algebraic methods resulted in the determination of the
equation of motion of the Green functions \cite{BrouderEuroLett}
and the description of the relations between general and connected Green functions
in the presence of a general state \cite{BrouderMN}.

The relation between connected and one-particle irreducible
(1PI) Green functions, which is the main topic of the present article, is more difficult to understand and depends
on the very definition of when a diagram is irreducible.
Here, we show that a rather natural definition
leads to a complete description of connected Green functions
in terms of 1PI Green functions.

We should point out that our main long-term interest is in the study of the 
electronic structure of highly correlated materials by means of Green 
functions, also in the framework of classical (non-relativistic) quantum 
chemistry. However, we restrict here our study of one-particle-irreducibility 
to the particular case of local potentials. This covers the potentials of 
quantum field theories -including QED, which is meaningful for our long-term 
purposes-, but not the Coulomb potential. There are several ways to remedy 
this problem and extend the constructions in the present article to 
time-dependent perturbation theory in many-body physics. For example, as the 
referee pointed out, our results may be extended to the case of general 
interactions by means of the notion of block truncation as introduced 
in~\cite{Djah}. However, for the sake of simplicity -and since the 
combinatorics of one-particle-irreducibility is already intricate enough 
for local potentials, we decided to postpone the study of electronic systems 
(and of non-scalar fields) to further work.

The paper starts with a short introduction to Green functions in the presence of initial correlations and to Hopf algebras,
followed by the definition of quantum field forms and
their convolution logarithm. Then, the relation between 
forms and connected forms is made explicit, providing
the classical relation between general Green functions
and connected Green functions. To discuss 1PI functions,
we need to generalize a recent work by Mestre and Oeckl
\cite{Mestre,Mestre2}. Then, a rather natural definition
of 1PI functions will be proposed and the Mestre-Oeckl
approach will be used to write a connected Green function
in terms of these 1PI functions. In the process, universal 
properties of symmetric functions with respect to Hopf algebra 
derivations are put forward.

\section{Green functions with initial correlations}
In the present section, we fix the notation and briefly survey the 
definition of Green functions, emphasizing the role of the initial state.

\subsection{Field operators}
Quantum fields are operator-valued distributions acting on a Fock space
\cite{ReedSimonII}. 
Here, we describe the construction of Fock
space, creation and annihilation operators and the corresponding
quantum fields. 
We start from a self-adjoint operator $h$ acting on
a Hilbert space $\mathcalH$ and, for notational convenience,
we assume that $h$ has a pure point spectrum, so that
there is an orthonormal basis $\ket{e_i}$ (with $i\in I$)
of $\mathcalH$ consisting of eigenvectors of $h$.
In many applications, the Hilbert space $\mathcalH$ is a function space
and the eigenvectors (written then preferably with the functional notation $\phi_n$) are functions of $\bfr$
(where $\bfr$ is a point in three-dimensional space).

The tensor product of Hilbert spaces is well defined
(see Ref.~\cite{ReedSimonI} p.~49) and the
\emph{symmetric Fock space} over $\mathcalH$ is 
the Hilbert space $S(\mathcalH)=\bigoplus_{N=0}^\infty S^N(\mathcalH)$,
where $S^N(\mathcalH)$ is the $N$-fold symmetric tensor product
of $\mathcalH$. An orthogonal basis of the vector space $S^N(\mathcalH)$ is 
provided by the set of vectors
\begin{eqnarray}
\ket{e_{i_1}}\vee \dots \vee \ket{e_{i_N}} &=& \frac{1}{\sqrt{N!}}
\sum_{\sigma} \ket{e_{i_{\sigma(1)}}}\otimes \dots\otimes 
    \ket{e_{i_{\sigma(N)}}},
\label{ei1eiN}
\end{eqnarray}
where $\sigma$ runs over the permutations of
$N$ elements and 
where $(i_1,\dots,i_N)$ runs over the subset of
$I^N$ such that $i_1\le\dots\le i_N$.
In this formula, the symbol $\vee$ denotes the symmetric product
and $1/\sqrt{N!}$ is a normalization factor.

In many-body theory, $S^N(\mathcalH)$ is called the $N$-particle space
of the system and its elements are the $N$-particle states.
In particular, $S^0(\mathcalH)$ is a one-dimensional vector space
denoted by $\mathbb{C}1$ in the mathematical literature.
In many-body physics and quantum field theory, this unit 1 of the 
tensor product is denoted by $\ket{0}$, this is the vacuum of the theory
(i.e. the state without a particle).

The creation operator $a^\dagger_n$ is defined as the linear
map from $S(\mathcalH)$ to itself such that, for any 
basis vector $\ket{u}$ of $S(\mathcalH)$, 
$a^\dagger_n \ket{u}=c_n(u) \, \ket{e_n}\vee \ket{u}$, where
$c_n(u)$ is a normalization factor (see for example \cite{Fetter}).
It is called a creation operator because
it maps $S^N(\mathcalH)$ to $S^{N+1}(\mathcalH)$: it adds a new particle
to a $N$-particle state. Its adjoint $a_n$ is called
an annihilation operator. The normalization factor
ensures that the commutation relation
$a_m \opro  a^\dagger_n - a^\dagger_n \opro a_m=\delta_{nm}$
holds, where $\opro$ denotes the composition of operators.

In functional notation, the corresponding quantum field is the 
(self-adjoint) operator-valued distribution on the three-dimensional space:
\begin{eqnarray*}
\phiS(\bfr) &=& \sum_{n\in I} \phi_n(\bfr) a_n
+ \phi_n^*(\bfr) a^\dagger_n.
\end{eqnarray*}
This formalism is used to describe scalar particles or photons
(up to an additional vector index in that case;
recall that a self-adjoint field operator
describes a neutral particle, charged scalar or fermion field
operators are not self-adjoint). We remind that we focus in the present article on scalar particles (without a charge).

\subsection{Adiabatic limit}
The adiabatic limit is a very general way of solving
the Schr\"odinger equation for a system described by
the Hamiltonian $H=H_0+V$ where the eigenstates
of $H_0$ are known but not those of $H$. 
The general idea behind the technique is that, for a particle evolving in a potential $V$, the effect of the potential on the motion can be treated (or can be expected to be treated) perturbatively. As we mentioned, besides the perturbative expansions of QFT, the same general idea shows up in the perturbation series of statistical physics \cite{Djah}.

As far as adiabatic limits are concerned, the basic idea is quite simple. We define a time-dependent 
Hamiltonian $H(t)=H_0 + \ee^{-\epsilon |t|} V$.
When $\epsilon$ is small, the interaction $H(t)$ 
is very slowly switched on from
$t=-\infty$ where $H(-\infty)=H_0$ to $t=0$
where $H(0)=H$. It is hoped that, if $\epsilon$
is small enough, then an eigenstate of $H_0$
is transformed into an eigenstate of $H$.

To implement this picture, the time-dependent
Schr\"odinger equation given by
$i\partial \ket{\Psi_S(t)}/\partial t=H(t) \ket{\Psi_S(t)}$
is solved.
However, the solution $\ket{\Psi_S(t)}$ is not convenient because
it has no limit when $t\to-\infty$. Therefore,
we define $\ket{\Psi(t)}=\ee^{i H_0 t} \ket{\Psi_S(t)}$ that satisfies
$i\partial \ket{\Psi(t)}/\partial t=\Hint(t) \ket{\Psi(t)}$ with respect to
$\Hint(t)=\ee^{iH_0t} V \ee^{-iH_0t} \ee^{-\epsilon|t|}$.
Now $\Hint(-\infty)=0$ and $\ket{\Psi(-\infty)}$ makes sense.
Using $\Hint$, we can start from the ground state $\ket{\Phi_0}$
of $H_0$ and solve the time-dependent 
Schr\"odinger equation with the boundary condition
$\ket{\Psi(-\infty)}=\ket{\Phi_0}$. When no eigenvalue crossing takes place,
$\ket{\Phi_0}$ should be transformed into the ground state
$\ket{\Psi(0)}$ of $H$.

Instead of calculating directly $\ket{\Psi(t)}$ it is 
convenient to define the unitary operator
$U(t)$ as the solution of
$i\partial U(t)/\partial t=\Hint(t) U(t)$, with the boundary
condition $U(-\infty)=1$. Thus, $\ket{\Psi(t)}=U(t)\ket{\Phi_0}$.
Note that $U(t)$ depends on $\epsilon$, as $\Hint(t)$.
But is $\lim_{\epsilon\to 0} U(0)\ket{\Phi_0}$ an eigenstate of $H$?
It would if the limit existed, but it does not~\cite{BSP}.
However, Gell-Mann and Low \cite{GellMann} discovered in 1951 that
\begin{eqnarray*}
\ket{\PsiGL} &=& \lim_{\epsilon\to0}\frac{U(0)\ket{\Phi_0}}
   {\bra{\Phi_0} U(0)\ket{\Phi_0}}
\end{eqnarray*}
exists and is an eigenstate of $H$. A mathematical proof
of this fact for reasonable Hamiltonians came much later
\cite{Nenciu}. Notice that
the above scheme works when the ground state of $H_0$
is non degenerate. When it is degenerate, the problem
is more subtle \cite{BP09,BMP10} and the limit $\epsilon\to0$ only exists
when $\ket{\Phi_0}$ is properly chosen \cite{BPS,BPS-PRL}.

\subsection{Green functions}
\label{sectGreen}
We now come to the heart of QFT: the calculation
of Green functions (or moment functions). Green functions are important because
they allow for the calculation of practically all relevant
physical observables: energy, charge density, transport 
coefficients, current density, dielectric constants, etc.
In particular, they show up naturally in the perturbative expansions arising from adiabatic limits.
If we could calculate Green functions exactly, we would
know all interesting properties of matter. 
Of course, as far as many-body theory is concerned, we cannot calculate exact Green functions
for realistic materials, but non-perturbative
approximations are now used with great success~\cite{Onida}.

When the dynamics of the particles is described by
a one-body Hamiltonian $H_0$,
the $n$-point Green function for scalar particles is
defined by
\begin{eqnarray*}
\Gzero_n(x_1,\dots,x_n) &=& 
  \bra{\Phi_0} T\big(\varphi(x_1)\dots \varphi(x_n)\big) \ket{\Phi_0},
\end{eqnarray*}
where $x=(t,\bfr)$, $T$ is the time-ordering operator
and $\varphi(x)$ is related to
$\phiS(\bfr)$ by
\begin{eqnarray*}
\varphi(x) &=& \ee^{i H_0t} \phiS(\bfr)\ee^{-i H_0 t}
= \sum_{n\in I} \ee^{-i\epsilon_n t} \phi_n(\bfr) a_n
+ \ee^{i\epsilon_n t} \phi_n^*(\bfr) a^\dagger_n,
\end{eqnarray*}
where the $\phi_n(\bfr )$ are eigenvectors of $H_0$ with associated eigenvalues $\epsilon_n$.
The time-ordering operator orders the quantum fields
$\varphi(x_1),\dots,\varphi(x_n)$ so that the field
$\varphi(x_i)$ is on the left 
of $\varphi(x_j)$ if $t_i$ is greater (i.e. later) than
$t_j$. For example
$T\big(\varphi(x_1)\varphi(x_2)\big)=\varphi(x_1)\varphi(x_2)$
if $t_1>t_2$ and
$T\big(\varphi(x_1)\varphi(x_2)\big)=\varphi(x_2)\varphi(x_1)$
if $t_1<t_2$.

When the dynamics of the particles is described by 
a Hamiltonian $H=H_0+V$, where $H_0$ is one-body,
the expression for the Green function becomes~\cite{Gross,Fetter}
\begin{eqnarray*}
G_n(x_1,\dots,x_n) &=& \frac{
  \bra{\Phi_0}T\big( \varphi(x_1)\dots \varphi(x_n) 
    \ee^{-i\int \Hint(t) \dd t}\big)\ket{\Phi_0}}{
    \bra{\Phi_0}T\big(\ee^{-i\int \Hint(t) \dd t}\big)\ket{\Phi_0}},
\end{eqnarray*}
where 
$\Hint(t) = \ee^{iH_0t} V \ee^{-iH_0t} \ee^{-\epsilon|t|}$
and the limit $\epsilon\to0$ is implicitly taken.
This generalizes to non-scalar particles and, 
in the example of the non-relativistic electrons,
\begin{eqnarray*}
\Hint(t) &=& \ee^{-\epsilon|t|}
 \frac{1}{2}
   \int \dd \bfr \dd \bfr' \psi^\dagger(t,\bfr) \psi^\dagger(t,\bfr')
     V_{ee}(\bfr-\bfr') \psi(t,\bfr') \psi(t,\bfr).
\end{eqnarray*}

We are now ready to enter correlated systems. Assuming that
the initial state $\ket{\Phi_0}$ is the free field vacuum
$\ket{0}$ implies the classical expansion of Green functions 
in terms of Feynman propagators and, ultimately, of (usual) Feynman diagrams.
In many-body theory, the use of this decomposition of Green function
into sums of (usual) Feynman diagrams is restricted to very specific
states $\ket{\Phi_0}$ called \emph{quasi-free states}.
For the other states (or, equivalently, for the study of general functional measures in statistical physics), the structure of Green functions is
more complex.
Let us give a simple example.
We can define the quantity 
$D_4(x_1,x_2,x_3,x_4)$ by
\begin{eqnarray*}
\Gzero_4(x_1,x_2,x_3,x_4) &=&
\Gzero_2(x_1,x_2) \Gzero_2(x_3,x_4)+
\Gzero_2(x_1,x_3) \Gzero_2(x_2,x_4)\\ &&+
\Gzero_2(x_1,x_4) \Gzero_2(x_2,x_3)
+ D_4(x_1,x_2,x_3,x_4).
\end{eqnarray*}
When the initial state is the vacuum or a quasi-free state, the
term $D_4$ is zero. For a general initial state, it is not.

For a fermionic system, a term $D_4$ can be defined similarly. It
is absent when the ground state of $H_0$ can be written as
a Slater determinant. It is present when
the ground state of $H_0$ is degenerate,
as in open shell systems\footnote{For non-equilibrium systems,
additional complications come from the fact that
time-ordered products must be defined over a closed
time path \cite{Chou}. However, this does not change the
combinatorial aspects of the problem.}.
In that case, $\ket{\Phi_0}$ can be written as a linear combination of
Slater determinants and $D_4$ describes the correlation 
between these determinants. 
The presence of several Slater determinants in the initial
state is rather catastrophic for many-body theory.
Yaris and Taylor summarized the situation~\cite{YarisTaylor}:
``The inability to handle open-shell systems is a ubiquitous problem
in many-body theory. It basically arises when one cannot find a
single-determinant unperturbed ground state which connects to the
exact ground state when the residual interaction is adiabatically
switched on. When this situation holds, one cannot properly define
occupied and unoccupied single-particle states, Wick's theorem
does not hold, and Dyson equations, Bethe-Salpeter equations, etc.
do not exist.'' To this list one can add that ordinary Feynman diagrams
and the Gell-Mann and Low formula are lost.
In other words, most of the tools of
quantum field theory break down.
Since the seminal work by Bloch and Horowitz in 1958~\cite{BH58}, many
works were devoted to the rebuilding of these tools.
Morita discovered a modified version of the Gell-Mann and Low 
theorem~\cite{Morita}, Fujita defined generalized Feynman 
diagrams~\cite{Fujita}, 
Hall derived a Dyson equation for degenerate systems \cite{Hall}.
Since then, progress has been quite slow because of the combinatorial
complexity of the problem.

To illustrate this complexity, we first describe the generalized
Feynman diagrams introduced by Fujita~\cite{Fujita}, Hall~\cite{Hall}
and Djah et al.~\cite{Djah}.
For bosonic and fermionic systems, $D_4$ can be thought
of as a sort of 4-point Feynman propagator, 
as $D_2(x,y)=G^0_2(x,y)$ is the 2-point Feynman propagator.
We shall see that $D_4$ plays the role of a cumulant,
as in the decomposition of a distribution function. 
Higher order Green functions $G^0_{2n}$
give rise to higher order propagators $D_{2n}$
and the precise relation between them will be described in
the following. In standard quantum field theory, the 
Green function of the interacting system can be written
by adding all possible Feynman diagrams involving the 
two-point propagator $D_2$. When the initial state is not
quasi-free, the Green function is written as the sum of
all possible Feynman diagrams involving 2-point,
4-point, and $2n$-point propagators for arbitrary $n$.
An example will be given in figure \ref{figGreen} of this paper.

\section{Hopf algebra}
\label{Hopfalgsect}
We do not provide the general definition
of a Hopf algebra (see e.g. \cite{Majid}) and consider only the
special case of the symmetric Hopf algebra 
$S(V)=\bigoplus\limits_{n}S^n(V)=\bigoplus\limits_nV^{\otimes n}/\mathcal{S}_n$,
where $V$ is a complex vector space and where $\mathcal{S}_n$, the symmetric 
group of order $n$, acts by permutation on the components of the tensor 
power $V^{\otimes n}$.  The commutative product of $S(V)$ is denoted by
concatenation.  The counit is the linear map $\counit: S(V)\to\mathbb{C}$
defined by $\counit(1)=1$, $\counit(u)=0$ if $u\in S^n(V)$ with $n>0$.
The coproduct is the linear map
$\Delta: S(V) \to S(V)\otimes S(V)$ 
determined by $\Delta 1 = 1\otimes 1$,
$\Delta a = 1\otimes a + a \otimes 1$ for $a\in V$
and $\Delta (uv)=(\Delta u)(\Delta v)$, for 
$u$ and $v$ in $S(V)$.
We employ the strengthened Sweedler notation for the coproduct \cite{Swe:hopfalg}:
$\Delta u = u\ix1 \otimes u\ix2$. Recall that there is an implicit summation in the notation, which does not lead to ambiguities when handled correctly: the right hand side should be understood not as the mere tensor product of two elements in $S(V)$ but as a sum of such elements (so that e.g. $a\ix1\otimes a\ix2=1\otimes a + a \otimes 1$ for $a\in V$). More generally, an expression such as $u\ix1v\ix1\otimes u\ix2v\ix2$, which stands for $\Delta (u)\Delta(v)=\Delta(uv)$, contains an implicit double summation and should be understood as: $(u\ix1\otimes u\ix2)(v\ix1\otimes v\ix2)$, and similarly for expressions of higher orders.

The \emph{iterated coproducts} $\Delta^k$ are defined by 
$\Delta^0 =\id$, $\Delta^1 =\Delta $ and
$\Delta^{k+1}  = (\id^{\otimes k}\otimes\Delta)\Delta^k $.
Their action on an element $u$ of $S(V)$ is denoted by
$\Delta^k u = u\ix1\otimes\dots\otimes u\ii{k+1}$.
For any $u\in S(V)$, the \emph{reduced coproduct} is the map 
$\Deltau: S(V) \to S(V)\otimes S(V)$ 
such that $\Deltau u = \Delta u -1\otimes u -u\otimes 1$.
The iterated reduced coproducts $\Deltau^k$ are defined by 
$\Deltau^0 =\id$, $\Deltau^1 =\Deltau $ and
$\Deltau^{k+1}  = (\id^{\otimes k}\otimes\Deltau)\Deltau^k $.
Their action on an element $u$ of $S(V)$ is denoted by
$\Deltau^k u = u\iu1\otimes\dots\otimes u\iiu{k+1}$. 
The coproduct and the reduced coproduct are cocommutative, that is:
$$\Delta (u)=u\ix1\otimes u\ix2=u\ix2\otimes u\ix1,
\quad \Deltau (u)=u\iu1\otimes u\iu2=u\iu2\otimes u\iu1.$$

The coproduct is an algebra morphism, but the reduced
coproduct is not. Its relation with the product is
described by the following simple and useful lemma.

\begin{lem}
\label{lemDeltaukau}
If $a\in V$ and $u\in S(V)$, then 
\begin{eqnarray*}
\Deltau(au) &=& 
  a \otimes u + u \otimes a +  au\iu1\otimes u\iu2
   +  u\iu1\otimes a u\iu2,
\end{eqnarray*}
and, for $k>1$,
\begin{eqnarray*}
\Deltau^{k}(au) &=& a\otimes \Deltau^{k-1}u
  +  au\iu1\otimes \Deltau^{k-1}u\iu2
  +  u\iu1\otimes \Deltau^{k-1}(au\iu2).
\end{eqnarray*}
More explicitly, for $k>0$,
\begin{eqnarray*}
\Deltau^{k}(au) &=& \sum_{i=1}^{k+1}
  u\iu1\otimes \dots \otimes u\iiu{i-1}\otimes a \otimes u\iiu{i}\otimes\dots
  \otimes u\iiu{k}
\\&&
+ \sum_{i=1}^{k+1}
  u\iu1\otimes \dots \otimes u\iiu{i-1}\otimes au\iiu{i}
   \otimes u\iiu{i+1}\otimes\dots \otimes u\iiu{k+1},
\end{eqnarray*}
where the terms $i=1$ and $i=k+1$ are $a\otimes \Deltau^{k}u$
and $\Deltau^{k}u \otimes a$ in the first sum and
$(a\otimes 1^{\otimes k})\Deltau^{k}u$ and
$(1^{\otimes k}\otimes a)\Deltau^{k}u$ in the second term. 

For an arbitrary $u\in S^n(V), n>0$ and $v\in S(V)$, we also have:
\begin{eqnarray*}
\Delta^k(uv) &=& u\ix1v\ix1\otimes \dots\otimes u\ix{k}v\ix{k}\\
&=& \sum_{1\leq p\leq k}\sum_{1\leq i_1<\dots<i_p\leq k} 
v\ix1\otimes\dots\otimes u\iu{1}v\ix{i_1}\otimes\dots\otimes 
u\iiu{p}v\ix{i_p}\otimes\dots\otimes v\ix{k}.
\end{eqnarray*}
\end{lem}

\section{Green functions for quasi-free states}
Let $V$ be the vector space generated by the symbols
$\varphi(x)$, where $x$ runs over points of $\mathbb{R}^d$.
In physical terms, $\varphi$ should be thought of as a 
free bosonic field operator, that is, as an operator-valued distribution 
(think of the quantum fields $\phi_{\bf S}(\bfr)$). Our forthcoming 
developments can be adapted easily to fermionic systems, the adaptation 
amounting mathematically to replacing the symmetric algebra $S(V)$ by 
the exterior (or Grassmann) algebra $\Lambda(V)$, see \cite{BrouderQG}. 

Defining a time-ordered product of fields at the same 
point gives rise to major difficulties and is the subject 
of renormalization \cite{Collins}. 
Here, we take advantage of the fact that the combinatorics of Green functions 
is in many respects a self-contained topic and leave aside these questions 
(renormalization, operator product expansion). We will therefore treat powers 
of fields such as $\varphi^4(x)$ as formal expressions, that is as monomials 
belonging to the symmetric Hopf algebra $S(V)\supset S^4(V)$.
Note that $\varphi^0(x)=1$ is the unit of the algebra 
$S(V)$.\footnote{In another paper \cite{BrouderMN}, an algebra
 different from $S(V)$ was used, where $\varphi^0(x)$ was not
 the unit of the algebra, in order to obtain some desirable coalgebraic 
 properties. 
 That alternative construction considers the field products $\varphi^n(x)$ 
 as the basis of a Hopf algebraic fiber at $x$. However such a point of
 view is not required in the present paper.}

\subsection{Convolution}
In this section we survey some Hopf algebraic concepts that provide
a startling simplification of the decomposition of 
the expectation value of time-ordered products in terms of 
Feynman diagrams.
We define a \emph{form} as a linear map from
$S(V)$ to $\mathbb{C}$.
A \emph{unital form} is a form $\rho$ such that $\rho(1)=1$.
In our context, that is when $\varphi(x)$ is the quantum field of QFT or 
many-body theory, unital forms are defined from states of $H_0$:
if $\ket{\Phi}$ is a normalized state and $u\in S(V)$, then
$\rho(u)=\bra{\Phi} T(u) \ket{\Phi}$ is a unital form because
it is obviously linear and
$\rho(1)=\bra{\Phi} 1 \ket{\Phi}=1$.
The unital form corresponding to the vacuum is denoted by
$\rho_0$, so that $\rho_0(u)=\bra{0} T(u) \ket{0}$.

To express $\rho_0(u)$ in Hopf algebraic terms, we first
need a few definitions.
The \emph{convolution} product of two forms $\rho$ and $\sigma$
is the form $\rho \ast\sigma$ defined by
$(\rho \ast\sigma)(u)= \rho(u\ix1) \sigma(u\ix2)$. Notice that, because of the commutativity and cocommutativity of $S(V)$, $\sigma\ast \rho=\rho\ast \sigma$.
The space of unital forms equipped with the convolution product
is a commutative group, denoted by
$\mathcal S$, whose unit is the counit $\counit$.

The $n$-th \emph{convolution power} of a form $\rho$ is the form
$\rho^{\ast n}$ defined recursively by $\rho^{\ast 0}=\counit$,
$\rho^{\ast 1}=\rho$ and $\rho^{\ast(n+1)}=\rho^{\ast n}\ast\rho$.
The \emph{convolution exponential} of a form $\rho$
is the form $\ee^{\ast\rho}$ defined by
\begin{eqnarray*}
\ee^{\ast\rho} &=& \sum_{n=0}^\infty \frac{\rho^{\ast n}}{n!}.
\end{eqnarray*}
The convolution logarithm $\log^\ast\rho$ of the form $\rho$
is the form defined by
\begin{eqnarray*}
       \log^\ast\rho &=& \sum_{n=1}^\infty \frac{(-1)^{n+1}}{n}
       (\rho-\epsilon )^{\ast n}.
\end{eqnarray*}
Note that, if $\rho$ is a unital form, $\log^\ast\rho$ satisfies 
$\log^\ast\rho(1)=0$. A form $\sigma$ such that
$\sigma(1)=0$ is called an \emph{infinitesimal form}
because it is the logarithm of a unital form.
In the present paper, the convolution exponential 
is always applied to infinitesimal forms.
Note that, if $\sigma=\log^\ast\rho$, then
$\ee^{\ast\sigma}=\rho$. In other words, convolution exponential
and convolution logarithm are inverse functions of each other.
At last, note that, if $\alpha$ and $\beta$ are two unital forms with
convolution logarithms $a$ and $b$, then
$\alpha \ast \beta=\ee^{\ast a}\ast\ee^{\ast b}=\ee^{\ast(a+b)}$. 

\subsection{Expansion in Feynman diagrams}
In standard quantum field theory, Wick's theorem states that,
if $u=\varphi^{k_1}(x_1)...\varphi^{k_n}(x_n)$, $<0|T(u)|0>$ 
is calculated as the sum
of all pairings of $k_1$ times the point $x_1$,
\dots, $k_n$ times the point $x_n$. A pairing
is the choice of a pair of different points represented graphically as a line and
analytically as a Feynman propagator. 
Graphically, $\rho_0(u)=\bra{0}T(u)\ket{0}$ is therefore represented 
by the sum of all the
graphs with $n$ vertices labeled by $x_1,\dots,x_n$ such that
$k_i$ edges are incident to the vertex labeled by $x_i$,
for $i=1,\dots,n$. Each graph is weighted by a proper
combinatorial factor.

To express $\rho_0(u)$ in Hopf algebraic terms, we
define the infinitesimal form $\tau$ by
$$\tau (\varphi(x_1)\varphi(x_2)):=D_F(x_2-x_1) \ 
\mathrm{if}\ x_1\not=x_2,$$ and
$$\tau (\varphi(x_1)...\varphi(x_n)):=0,\ \mathrm{if}\ n\not= 2
\  \mathrm{or}\ n=2 \ \mathrm{and}\ x_1=x_2.$$
The form $\tau$ is called the \emph{Feynman form}. 
The function $D_F$ is defined\footnote{When $x_1$ and
$x_2$ are separated by a light-like interval, the definition
of $\tau$ does not make sense and $D_F$ should be replaced by a
smooth regularization. We do not enter into these details here
since we consider only the combinatorial aspects of the
problem.} by
$$D_F(x)=\int
\frac{d^4p}{(2\pi)^4}\frac{i}{p^2-m^2+i\varepsilon}e^{-i(p\cdot x)}.$$

We can now restate Wick's theorem algebraically:

\begin{thm}(\cite{Ruelle,Borchers,BrouderQG,BrouderMN}) \label{Wick} 
The unital form $\rho_0$ is the convolution exponential of the Feynman 
form : $$\rho_0 =e^{\ast \tau}.$$
\end{thm}

This theorem extends to the case of quasi-free states \cite{scutaru},
the only change is that $\tau$ is now defined by
$$\tau (\varphi(x_1)\varphi(x_2)):=D_2(x_1,x_2) \ 
\mathrm{if}\ x_1\not=x_2,$$ and
$$\tau (\varphi(x_1)...\varphi(x_n)):=0,\ \mathrm{if}\ n\not= 2
\  \mathrm{or}\ n=2 \ \mathrm{and}\ x_1=x_2,$$
where 
$D_2(x_1,x_2)=\bra{\Phi}  T\big(\varphi(x_1)\varphi(x_2)\big) \ket{\Phi}$.

\section{Green functions for general states}

Most often the relevant 
object to deal with in perturbative expansions is actually not the 
unital form $\rho$ built from the ground state $\ket{\Phi}$
(or, abstractly, the group $\mathcal S$) but its convolution 
logarithm $r$ (resp. the corresponding 
commutative Lie algebra $\mathcal L$).
The infinitesimal form $r$ is called the \emph{cumulant form}.
As we shall see in section \ref{phitrois}, this is exactly what we
need to calculate the Green functions in the presence of initial correlations.

The theorem \eqref{Wick} of the previous section trivially generalizes to arbitrary states and unital forms:
\begin{thm}(\cite{Ruelle,Borchers,Djah}) \label{Wickgen} The unital form 
$\rho$ is the convolution exponential of its cumulant
form : $$\rho =e^{\ast r}.$$
\end{thm}

Although our approach is not the usual one, writing $\rho$ as 
$\ee^{\ast r}$ is in fact quite common in physics.
The notion of a cumulant form is related to the cumulant expansion, 
and expresses the
generalized Wick theorem used for solving the Anderson model
\cite{Moskalenko}. Moreover, it is a way to isolate the
singularities of the forms because a natural property of
a quantum field is that $r(\varphi(x_1)\dots\varphi(x_n))$
is a smooth function of $x_1,\dots,x_n$, except possibly for $n=2$ 
(see \cite{Hollands}, as well as \cite{TikhodeevCor} for a related result 
in many-body theory).

\begin{figure}[!ht]
\begin{center}
\includegraphics[width=4cm]{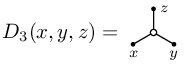}
\end{center}
\caption{The generalized propagator
  $D_3(x,y,z)=r\big(\varphi(x)\varphi(y)\varphi(z)\big)$ \label{propag}}
\end{figure}

Finally, an observation that will prove essential in our forthcoming 
developments: all our previous reasonings suggest that a unital form 
$\rho=\ee^{\ast r}$ 
should be dealt with by means of
generalized propagators in the same way as vacuum expectations of 
time-ordered products of free fields are dealt with by means of 
2-point Green functions and Feynman propagators in the usual picture of QFT. However, whereas the 
Feynman propagator, which is associated to the unique non trivial component 
of $\tau $ (recall that $\tau =0$ on $S^i(V)$ for any $i\not= 2$), is
described graphically by a line linking two vertices, 
we may have now $r(\varphi(x_1)\dots
\varphi(x_n))\not=0$ with $n\not= 2$. Accordingly, we shall represent 
graphically the ``$n$-point propagator'' 
$D_n(x_1,\dots,x_n)=r(\varphi(x_1)\dots\varphi(x_n))$\footnote{The 
definition of $D_n$ generalizes the definition of $D_4$ in 
section~\ref{sectGreen} --this should be clear from our forthcoming
developments.}
by a white dot with $n$ edges linked to the $n$ vertices $x_1,...,x_n$,
as shown in figure \ref{propag} (a similar convention was
used by Djah et al. \cite{Djah}).

As we already mentioned, we can also consider $\rho=\ee^{\ast r}$ as 
a generalization of Wick's theorem when the latter is stated algebraically. 
The same observation holds for graphical statements of the theorem:
we saw that, in standard quantum field theory, Wick's theorem states that,
if $u=\varphi^{k_1}(x_1)...\varphi^{k_n}(x_n)$, 
$\bra{0}T(u)\ket{0}$ is calculated as the sum
of all pairings of $k_1$ times the point $x_1,\dots,k_n$ times the point $x_n$. 
In the many-body context $\ee^{*\tau}(u)$ is replaced by
$\ee^{*r}(u)$. This amounts to say that we write 
$\ee^{*r}(u)$ as the sum of all ways to partition 
the multiset made of $k_1$ times point $x_1$,
\dots, $k_n$ times point $x_n$ into sub-multisets of any
multiplicity (i.e. not only pairs and not only different points). 
See figure~\ref{figGreen} for an example. The $n$-point propagators are then a 
convenient way to represent these sub-multisets.

To conclude this section, we state three easy but important lemmas
\begin{lem}
If $a\in V$ and $E=\ee^{ a}$, then for any form $\rho$
with logarithm $r$, we have $\rho(E) = \ee^{r(E)}$.
\end{lem}
\begin{lem}
\label{expraulem}
If $a\in V$ and $u\in S(V)$,
then, for any linear map $r: S(V)\rightarrow \mathbb{C}$
such that $r(1)=0$,
\begin{eqnarray*}
\ee^{*r}(au) &=& \sum r(au\ix1)\ee^{*r}(u\ix2).
\end{eqnarray*}
\end{lem}
More generally,
\begin{lem}
\label{lemrhouv}
For any $u\in\ker\epsilon$ and any $v$ in $S(V)$,
\begin{eqnarray}
\ee^{*r}(uv) &=& \sum_{k=1}^\infty \frac{1}{k!}
  r(u\iu1 v\ix1)\dots r(u\iiu{k}v\ii{k})\ee^{*r}(v\ii{k+1}).
\label{rhouvexp}
\end{eqnarray}

\end{lem}
\begin{proof}
The first lemma is a simple consequence of the fact that $E$
is group-like (that is, $\Delta(E)=E\otimes E$). The second lemma was
shown
in \cite{BrouderMN}, it follows from the cocommutativity of the
coproduct and
the fact that $a$ is a primitive element (that is, $\Delta (a)=1\otimes
a +
a\otimes 1$).
The third lemma follows from the last identity in
Lemma~\ref{lemDeltaukau},
from the properties of the binomial coefficients, and from the
cocommutativity of the coproduct.
\end{proof}
The first lemma is often used with $a=\int j(x) \varphi(x)\dd x$
(up to a suitable extension of the definition of $V$ when the function
$j(x)$ has not a discrete support).
In that case, it relates the generating function of the moments
of $\rho$ to that of the moments of $r$. The second and third lemmas
provide powerful tools for the recursive proof of the properties
of $\ee^{*r}$.
Notice in particular that, using the last Lemma with $v=1$:
\begin{eqnarray}
\rho(a^n) &=& \sum_{k=1}^n \frac{1}{k!} 
   \sum_{i_1+\dots+i_k=n} 
     \frac{n!}{i_1!\dots i_k!}
     r(a^{i_1})\dots r(a^{i_k}),
\label{rho(an)}
\end{eqnarray}
where, for $p=1,\dots,k$,
$i_p>0$. A formula with less terms can be given
using the Fa\`a di Bruno coefficients:
\begin{eqnarray}
\rho(a^n) &=& \sum_\alpha
   \frac{n! r(a^1)^{\alpha_1} \dots r(a^n)^{\alpha_n}}
{\alpha_1!(1!)^{\alpha_1}
\alpha_2!(2!)^{\alpha_2}\dots \alpha_n!(n!)^{\alpha_n}},
\label{Faa}
\end{eqnarray}
where $(\alpha_1,\dots,\alpha_n)$ are nonnegative integers
such that $\sum_i i\alpha_i=n$.
For the partition represented by $\alpha$, $n$ is cut
into $k=\sum_i \alpha_i$ parts.
For example, $\rho(a)=r(a)$, $\rho(a^2)=r(a^2) + r(a)^2$,
$\rho(a^3)=r(a^3) + 3r(a)r(a^2) + r(a)^3$.

%%%%%%%%%%%%%%%%%%%%%%%%%%%%%%%%%%%%%%%%%%%%%%%%%%%%%%%%%%%%%%%%%%%%%%%%%%%%%
%New Section
%%%%%%%%%%%%%%%%%%%%%%%%%%%%%%%%%%%%%%%%%%%%%%%%%%%%%%%%%%%%%%%%%%%%%%%%%%%%%

\section{Connected forms}
In quantum field theory, an important simplification
comes from the fact that all physical quantities
can be expressed in terms of connected diagrams.
We first define the notion of a \emph{connected form}
by analogy with that of a connected diagram.
A monomial of $S(V)$ can always be written
$u=\varphi^{n_1}(x_1)\dots \varphi^{n_k}(x_k)$,
where all points $x_i$ are distinct.
Now, for any form $\rho$ with convolution logarithm $r$, 
we use eq.~(\ref{Faa}) to expand $\rho(u)$ in terms of $r$.

\begin{prop} We have
\label{proptuples}
\begin{eqnarray*}
\rho(u) = \ee^{*r}(u) &=& \sum\limits_{l\in\N}\frac{1}{l!}
\sum\limits_{\substack{n_i^{1}+...+n_i^{l}=n_i\\i=1...k}}  
\prod\limits_{i=1}^k  \frac{n_i!}{n_i^{1}!\dots{n_i^{l}!}}
\\&& \times\, r(\varphi^{n_1^{1}}(x_1)
\dots \varphi^{n_k^{1}}(x_k))...r(\varphi^{n_1^{l}}(x_1)\dots 
\varphi^{n_k^{l}}(x_k)),
\end{eqnarray*}
where, for $i=1,\dots,k$, the sum is over all the $l$-tuples 
of nonnegative integers $(n_i^1,\dots,n_i^l)$ such that
$n_i^1+\dots+n_i^l=n_i$.\end{prop}
Although a straightforward application of the Hopf algebra formalism, the 
result is important since it allows us to compute the multiplicity of a 
graph --or symmetry factor-- in the Feynman diagrammatic perturbative 
expansion of amplitudes. We refer for example to the expansion of the 
connected Green functions for $\varphi^3$ theory with an arbitrary ground 
state in the present section of the article.

Let us consider a term $t:=r(\varphi^{n_1^{1}}(x_1)
\dots \varphi^{n_k^{1}}(x_k))\dots r(\varphi^{n_1^{l}}(x_1)\dots 
\varphi^{n_k^{l}}(x_k))$ of $\ee^{*r}(u)$. We say that 
$x_i\cong_t x_j, 1\leq i,j\leq k$ if there exists $m\leq l$ with 
$n_i^{m} n_j^{m}\not=0$. The transitive closure $\equiv_t$ of 
the binary relation $\cong_t$ defines the connectedness of $t$: $t$ is said 
to have $n$ connected components if there are $n$ equivalence classes 
associated to the equivalence relation $\equiv_t$. The connected component 
of $x_i$ in $t$ is defined similarly as the product of all the 
$r(\varphi^{n_1^{m}}(x_1) \dots \varphi^{n_k^{m}}(x_k))$ with 
$n_j^{m}\not= 0$ for at least one coefficient $j$ with $x_i\equiv_t x_j$. 
When $n=1$ (resp. $n\not= 1$), we also say that the term $t$ is connected 
(resp. disconnected).  
Let us take a simple example. For $u=\varphi(x)\varphi^2(y)$, we have
$\rho(u)=r(\varphi(x))r(\varphi^2(y))
+ r(\varphi(x))r(\varphi(y))^2
+ 2 r(\varphi(x)\varphi(y))r(\varphi(y))
+ r(\varphi(x)\varphi^2(y))$,
where the first two terms are disconnected (they actually have two connected 
components). The connected components of $y$ in the four terms are 
respectively $r(\varphi^2(y)),r(\varphi(y))^2,r(\varphi(x)\varphi(y))
r(\varphi(y))$ and $r(\varphi(x)\varphi^2(y))$.
The definition of connected form is actually best formulated in 
algebraic terms: this is the purpose of the next section.

\subsection{Another coproduct on $S(V)$}
As we have just seen, a pedestrian
definition of connectedness makes an
essential use of the fact that some points
$x_i$ are equal or distinct. In order to reflect this distinction, we define a new coproduct, the
\emph{disconnecting coproduct} $\Deltad: S(V) \to S(V)\otimes S(V)$.

So we write a monomial of $S(V)$ as
$u=\varphi^{n_1}(x_1)\dots \varphi^{n_k}(x_k)$,
where all points $x_i$ are distinct,
and we define the coproduct of $u$ as follows:
$\Deltad \varphi^n(x) = 1\otimes \varphi^n(x) + \varphi^n(x)\otimes 1$
if $k=1$, 
and 
$\Deltad u = \Deltad(\varphi^{n_1}(x_1))\Deltad(\varphi^{n_2}(x_2))\dots
   \Deltad (\varphi^{n_k}(x_k))$ if $k>1$.
Notice that this coproduct is coassociative and cocommutative but is not an 
algebra morphism, because
$\Deltad(\varphi^2(x))\not= (\Deltad(\varphi(x)))^2$. Since $\Deltad$ is 
coassociative and cocommutative, we may still define an associative, 
commutative and unital product $\hat\ast$, the 
\emph{disconnecting convolution product}, on $Lin(S(V),\C )$:
$$\forall (f,g)\in Lin(S(V),\C), \ f\hat\ast g:=\pi\circ (f\otimes g)\circ 
\Deltad ,$$ where $\pi$ denotes the product:
$\pi\circ(f\otimes g)(u\otimes v)=f(u)g(v)$.
The unit of $\hat\ast $ is the same as the unit of $\ast$ (the projection 
map $\varepsilon$ from $S(V)$ to $\C\subset S(V)$). To distinguish between 
the two products $\ast$ and $\hat\ast$, we write the operations involving 
$\hat\ast$ with a superscript $\hat\ast$: for example, we write 
$\log^{\hat\ast}$, and so on.

The relation between $\Deltad$ and $\Delta$ is
investigated in \cite{BrouderMN}.
The reduced coproduct $\Deltadu$ and the iterated
coproduct $\Deltad^k$ are defined as in section \ref{Hopfalgsect}.
The enhanced Sweedler notation for the disconnecting coproduct is
$\Deltad u =  u\is1 \otimes u\is2$
and
$\Deltadu u =  u\isu1 \otimes u\isu2$.

The new coproduct $\Deltad$ enables us to give an
algebraic definition of the connected form
$\rhoc$ corresponding to the unital form $\rho$:
$$\rhoc =\log^{\hat\ast}(\rho )$$
that is,
       $$\forall u\in S(V),\ \rhoc(u)=\sum_{n=1}^\infty \frac{(-1)^{n+1}}{n}
       (\rho-\counit )(u\is1) {\dots} (\rho -\counit )(u\is{n}).$$
This identity can be understood as a linked cluster theorem for forms.
The same problem was addressed in \cite[Sect.6]{Djah} with another
approach: in that article, the authors introduce two notions of truncated moment functions, indicated respectively by an exponent $T$ and $(T)$. This corresponds roughly, at the Hopf algebraic level, to our distinction between the two coproducts $\Delta$ and $\Deltad$. We thank the referee for pointing out to us this point and refer to \cite{Djah} for further insights on truncated moment functions.

Pedantically, the set of connected forms is defined as the image of 
the group of unital forms under the map $\log^{\hat\ast}$. The two sets 
are in bijective correspondence and, reciprocally, we can express 
any unital form $\rho$ in terms of $\rhoc$ by 
$$\rho =\ee^{\hat\ast\rhoc}$$ or:
   $$\forall u\in S(V),   \rho(u) =\counit(u)+
       \sum_{n=1}^\infty \frac{1}{n!} \rhoc(u\is1)\dots \rhoc(u\is{n}).$$
For example, 
$\rhoc(\varphi^n(x))=\rho(\varphi^n(x))$ for $n>0$
and
$$\rhoc(\varphi^m(x)\varphi^n(y))=\rho(\varphi^m(x)\varphi^n(y))-
\rho(\varphi^m(x))\rho(\varphi^n(y))$$
for $m>0$ and $n>0$.
The connected form $\rhoc$ is an infinitesimal form
(that is, $\rhoc(1)=0$). For $u$ as above, $\rhoc(u)$ is defined as 
the sum of the connected terms of $\rho(u)$.
This terminology is due to the fact that we
can define Feynman diagrams to represent $\rho(u)$,
and that $\rhoc(u)$ is obtained by summing the connected
Feynman diagrams present in $\rho(u)$.

Note that the relation $\rho=\ee^{\hat\ast\rhoc}$ is the
analogue of the relation $Z=\ee^{W}$ between
the partition function and the free energy.
As we prove now, the latter is a consequence of the
former. The Hamiltonian density of a quantum field
theory of local interactions is 
of the form $H(x)$. Therefore, 
$\delta H(x)=H(x)\otimes 1 + 1 \otimes H(x)$ is primitive and, by lemma 5.1,
$Z = \rho(S) = \ee^{W}$, where $W=\rhoc(S)$
(see \cite{BrouderMN} for a detailed proof).
This is an extension of the standard relation
to the case of a general initial state.
As a matter of fact, the coproduct $\delta$ was
precisely defined for the Hamiltonian to be a primitive element.
This ensures the standard relation between the partition function
and the free energy (compare to \cite{Djah}).

\subsection{Example of the $\varphi^3$ theory}
\label{phitrois}
As we saw in section \ref{sectGreen},
the two-point Green function for a system described by
the interaction Hamiltonian density $u=\varphi^3(x)$ is given 
by the expression:
$$G(x,y)=
\frac{\langle 0| T(\varphi(x)\varphi(y)\ee^{-iu})|0\rangle}
{\langle 0| T(\ee^{-iu})|0\rangle}.$$
We recall that the denominator cancels the 
divergence of the adiabatic switching of the interaction.
In graphical terms, the denominator 
$\langle 0|T(\ee^{-iu})|0\rangle$ is the sum of all
the vacuum Feynman diagrams (i.e. the diagrams
that are linked neither to $x$ nor to $y$).
Another way to obtain a convergent expression is to
use the connected Green function $G_c(x,y)$ which is the sum of
all the connected diagrams in $G(x,y)$.

For a general form, the factorization of the adiabatic
divergence is more complex \cite{Morita,Kuo,Kitanin} and 
it holds only for specific initial states~\cite{BPS-PRL}.
For notational convenience, we do not write the denominator
in the definition of the Green functions for a general form
and we put
\begin{eqnarray}
G(x,y)=\rho\big(\varphi(x)\varphi(y)
   \ee^{-i\int_{-\infty}^\infty \Hint(t) }\big).
\label{Ggs}
\end{eqnarray}
The connected Green function is defined as
$$G_c(x,y)=\rhoc\big(\varphi(x)\varphi(y)
\ee^{-i\int_{-\infty}^\infty \Hint(t)}\big).$$

The term $\int_{-\infty}^\infty \Hint(t)$ can usually
be written $\int \dd x P(x)$, where $P(x)$ is 
a polynomial in $\varphi(x)$. Therefore,
\begin{eqnarray*}
G(x,y)=\rho\big(\varphi(x)\varphi(y)\big)
  + \sum_{n=1}^\infty \frac{(-i)^n}{n!}
   \int \dd x_1 \dots \dd x_n
\rho\big(\varphi(x)\varphi(y)
    P(x_1)\dots P(x_n)\big).
\end{eqnarray*}

For example, in the $\varphi^3$ theory,
we have $P(x)=\varphi^3(x)/3!$ and the
first terms of the total Green function are
\begin{eqnarray*}
G(x,y) &=& \rho\big(\varphi(x)\varphi(y)\big) 
  -\frac{i}{6} \int\dd x_1 \rho\big(\varphi(x)\varphi(y)\varphi^3(x_1)\big)
\\ &&  -\frac{1}{72} \int \dd x_1 \dd x_2
  \rho\big(\varphi(x)\varphi(y)\varphi^3(x_1)\varphi^3(x_2)\big)
+ \dots
\end{eqnarray*}
For notational convenience, we assume that 
$D_n(x_1,\dots,x_n)=r(\varphi(x_1)\dots\varphi(x_n))=0$ if $n$ is odd.
The connected Green function $G_c(x,y)$ is obtained by
keeping the connected terms of the total Green function.

In standard quantum field theory, the expansion to the
second order gives us
\begin{eqnarray}
G_c(x,y) &=& \propagator_2(x,y) - \frac{1}{2}\int \dd z\ \dd w\ 
\propagator_2(x,z)\ \propagator_2(y,w)\ \propagator_2(z,w)^2 + \dots 
\label{Gfacile}
\end{eqnarray}

\begin{figure}[!ht]
\includegraphics[width=12cm]{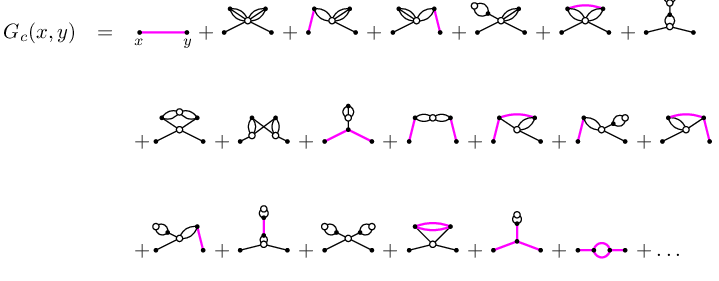}
\caption{The first few terms of $G_c(x,y)$ \label{figGreen}}
\end{figure}

For a general form, the expansion to the second order
gives a more complex result:
\begin{align}
G_c(x,y) &= \propagator_2(x,y) - \int \dd z\ \dd w\ 
\Big(\frac{1}{72}\propagator_8(x,y,z,z,z,w,w,w) \\ 
& \qquad + 
\frac{1}{12}\propagator_2(x,z)\ \propagator_6(y,z,z,w,w,w) 
\nonumber\\&
+ \frac{1}{12}\propagator_6(x,z,z,w,w,w)\ \propagator_2(y,z) 
+ \frac{1}{12}\propagator_6(x,y,z,w,w,w)\ \propagator_2(z,z) 
\nonumber\\&
+ \frac{1}{8}\propagator_6(x,y,z,z,w,w)\ \propagator_2(z,w) 
+ \frac{1}{12}\propagator_4(x,y,z,z)\ \propagator_4(z,w,w,w) 
\nonumber\\&
+ \frac{1}{8}\propagator_4(x,y,z,w)\ \propagator_4(z,z,w,w)
+ \frac{1}{4}\propagator_4(x,z,z,w)\ \propagator_4(y,z,w,w) 
\nonumber\\&
+\frac{1}{6}\propagator_2(x,z)\ \propagator_2(y,z)\ \propagator_4(z,w,w,w) 
+\frac{1}{4}\propagator_2(x,z)\ \propagator_2(y,w)\ \propagator_4(z,z,w,w)
\nonumber\\&
+\frac{1}{2}\propagator_2(x,z)\ \propagator_4(y,z,w,w)\ \propagator_2(z,w) 
+\frac{1}{4}\propagator_2(x,z)\ \propagator_4(y,z,z,w)\ \propagator_2(w,w)
\nonumber\\&
+\frac{1}{2}\propagator_4(x,z,w,w)\ \propagator_2(y,z)\ \propagator_2(z,w)
+\frac{1}{4}\propagator_4(x,z,z,w)\ \propagator_2(y,z)\ \propagator_2(w,w)
\nonumber\\&
+\frac{1}{4}\propagator_4(x,y,w,w)\ \propagator_2(z,z)\ \propagator_2(z,w) 
+\frac{1}{8}\propagator_4(x,y,z,w)\ \propagator_2(z,z)\ \propagator_2(w,w) 
\nonumber\\&
+\frac{1}{4}\propagator_4(x,y,z,w)\ \propagator_2(z,w)^2 
+\frac{1}{2}\propagator_2(x,z)\ \propagator_2(y,z)\ \propagator_2(z,w)\ 
  \propagator_2(w,w)
\nonumber\\&
+\frac{1}{2}\propagator_2(x,z)\ \propagator_2(y,w)\ 
     \propagator_2(z,w)^2\Big) + \dots 
\label{Gdur}
\end{align}
These terms can be given the diagrammatic representation of 
figure~\ref{figGreen}.

It is clear from a comparison of equations \eqref{Gfacile}
and \eqref{Gdur} that the use of a form $\rho$ which does not
come from a quasi-free state
increases significantly the combinatorial complexity. As a consequence, 
little was known about the structure of the
connected Green functions in the general case, especially as far as one-particle-irreducibility is concerned. 

%%%%%%%%%%%%%%%%%%%%%%%%%%%%%%%%%%%%%%%%%%%%%%%%%%%%%%%%%%%%%%%%%%%%%%%%%%%%%%%
%New Section
%%%%%%%%%%%%%%%%%%%%%%%%%%%%%%%%%%%%%%%%%%%%%%%%%%%%%%%%%%%%%%%%%%%%%%%%%%%%%%%
\section{Symmetric functions and derivations}

In \cite{Mestre,Mestre2}, Mestre and Oeckl have proposed a powerful
Hopf algebraic tool to generate 1PI diagrams. 
The forthcoming developments of the present article aim at
extending their work to correlated systems, and at describing in this general 
setting the decomposition of connected Green functions into
one-particle irreducible Green functions. 

In order to do so, the language of functional derivatives proves to be a very convenient framework. This leads naturally to the study of Hopf algebra derivations, which is the main topic of the present section.
Namely, we describe some aspects of the algebra of
symmetric functions in the Hopf algebraic setting and prove that this 
algebra has universal properties with respect to 
derivations acting on commutative Hopf algebras. We will show later that this 
formalism nicely encodes the properties of the generalized $n$-point
propagators associated to arbitrary states, as described above.
We do not seek the utmost generality in our constructions, but mention 
that they can be easily extended to more general (e.g. noncommutative) 
situations.

\subsection{Definition of the derivations $\deri_m$.}

Let us consider, once again, the Hopf algebra $U=S(V)$. 
Two products arise therefore in $S(U)=S(S(V))$: the symmetric product 
in $S(V)$, denoted by juxtaposition, 
and the symmetric product in $S(U)$, 
denoted from now on by $\vee$ to avoid any confusion.

To a unital form $\rho$ on $U=S(V)$, we associate its convolution
logarithm $r$ and, for $\varphi(x_1)\dots \varphi(x_m)
\in U$, we recall that
$D_m(x_1,\dots,x_m)=r(\varphi(x_1)\dots\varphi(x_m))$,
so that the generalized propagators $D_m$ are symmetric functions
of their arguments.

Recall that $V$ is spanned by the symbols $\varphi (x)$, where $x$ runs 
over points in $\R^d$. We choose an arbitrary total order on the points 
in $\R^d$, for example the lexicographical order on the $d$-tuples of 
coordinates.
The operators $\deri_{m-1}: U \to S^m(U)$  are then defined by
\begin{eqnarray}
\deri_{m-1}(u) &=& \sum\limits_{x_1<...<x_m} D_m(x_1,\dots,x_m)
  \frac{\overline\partial u\is1}{\partial \varphi(x_1)}\vee \dots \vee
  \frac{\overline\partial u\iis{m}}{\partial \varphi(x_m)}\nonumber\\
  &=&\frac{1}{m!}\,\sum\limits_{x_1,...,x_m} D_m(x_1,\dots,x_m)
  \frac{\overline\partial u\is1}{\partial \varphi(x_1)}\vee \dots \vee
  \frac{\overline\partial u\iis{m}}{\partial \varphi(x_m)},
\label{defAm}
\end{eqnarray}
where $\frac{\overline\partial u}{\partial \varphi(x)}$ is defined by the following (topologically motivated) rule:
\begin{itemize}
\item $\frac{\overline\partial u}{\partial \varphi(x)}:=\frac{\partial u}{\partial \varphi(x)}$ if $u$ is not linear in $\varphi(x)$ or if $\frac{\partial u}{\partial \varphi(x)}$ is a scalar.
\item $\frac{\overline\partial u}{\partial \varphi(x)}:= 0$ otherwise.
\end{itemize}
These operators on $U$ extend uniquely to derivations on $S(U)$, still written $\deri_{m-1}$, so that:
$$\deri_{m-1}: S^k(U)\longrightarrow S^{k+m-1}(U)$$
(recall, for completeness sake, that a derivation $D$ satisfies the Leibniz rule, $D(xy)=D(x)y+xD(y)$, so that the action of a derivation on $S(U)$ follows by induction on the degrees from the knowledge of its action on $U$).
Notice also that, by definition of $\delta$, the terms with $x_i= x_j$ 
vanish in the sum of eq.~(\ref{defAm}).

In practice, forcing the null value in the linear case (when $u$ is not a 
scalar multiple of $\varphi(x)$) will permit to avoid the creation of 
topologically disconnected graphs. We are particularly grateful to Achim 
Randelhoff who pointed out to us this property and allowed us to correct 
an error in a previous version of this article. The meaning of this 
observation should become clear later on, but the reader may immediately 
understand its implications by comparing the action of $\deri_{1}^2$ on 
$\varphi(x)\varphi(y)\varphi(z)^2$ with the action of the derivation 
$\hat\deri_{1}^2$, where the action of $\hat\deri_{m-1}$ is defined on $U$ by:
$$\hat\deri_{m-1}(u) = \sum\limits_{x_1<...<x_m} D_m(x_1,\dots,x_m)
  \frac{\partial u\is1}{\partial \varphi(x_1)}\vee \dots \vee
  \frac{\partial u\iis{m}}{\partial \varphi(x_m)}.$$

An important point for our forthcoming developments is that the $\deri_i$ 
commute. This follows from the Schwarz commutation rules 
for derivatives and from the definition of the disconnecting
coproduct $\delta$.
Graphically, these operations will allow us to construct inductively 1PI graphs (with a given set of vertices and with given multiplicities), and their algebraic properties will allow us to enumerate these graphs. This will be the purpose of the next section, whereas the following subsections introduce the abstract Hopf algebraic framework suited for this enumeration.

\subsection{The Hopf algebra of symmetric functions}

Let $X=\{x_1\dots x_n\dots \}$ be a countable alphabet, and $\CC [[X]]$ the 
algebra of formal power series over $X$. The group 
${\mathcal S}_\infty =\lim\limits_{\rightarrow}{\mathcal S}_n$ acts on $\CC [[X]]$ by 
permutation of the letters of $X$; the algebra of symmetric functions 
$\SYM$ is the subalgebra of ${\mathcal S}_\infty$-invariant series in $\CC [[X]]$. 

The algebra $\SYM$ is (up to completion with respect to the filtration 
induced by the grading of symmetric polynomials by their degree) a free 
commutative algebra over various families of generators. For our purposes, 
the most interesting ones are the families of power sums symmetric 
functions and complete symmetric functions associated respectively to 
the series
$${\mathbf P_\bullet}:=\sum\limits_{k\in\NN} P_k:=
1+\sum\limits_{i\in\NN}\frac{x_i}{1-x_i}$$
and
$${\mathbf C_\bullet}:=\sum\limits_{k\in\NN} S_k:=
\prod\limits_{i\in\NN}\frac{1}{1-x_i}.$$
In view of our forthcoming computations, it is actually convenient to work 
with an extension of $\SYM$, $q\SYM$: we write $Q_k$ for $\frac{P_k}{k}$, 
$k\geq 1$, $Q_0:=q$ and ${\mathbf Q_\bullet}:=\sum\limits_{k\in\NN} Q_k$, 
where $q$ stands for an additional free variable. The series are related by the familiar Newton-type identity:
${\mathbf C_\bullet}=e^{\mathbf Q_\bullet -Q_0}$. We write 
${\mathbf S}_\bullet$ for the $q$-series ${\mathbf S}_\bullet:=
e^{Q_0}\,{\mathbf C}_\bullet =e^{\mathbf Q_\bullet}$

The algebra $q\SYM$ carries a natural notion of grading (by the degrees of 
symmetric polynomials, with $deg(Q_n)=n$), but it is convenient, for our 
purposes, and for reasons that will become clear later, to introduce an
extra ``auxiliary'' grading by considering the family of the $Q_k$ as a 
family of generators of $\SYM$ of auxiliary degree 1. This is best 
explained through an example:
$Q_0^2Q_3Q_5Q_9$ is of degree $17$ and of auxiliary degree $5$. The 
auxiliary degree is indicated with a superscript (whereas the degree is 
indicated by a subscript), so that, for example, the component of degree 
$n$ and auxiliary degree $k$ in ${\mathbf S_\bullet}$ is given by:
 $${\mathbf S}_n^k=\sum\limits_\alpha\frac{Q_0^{\alpha_0}}{\alpha_0!}
\frac{Q_1^{\alpha_1}}{\alpha_1!}...\frac{Q_n^{\alpha_n}}{\alpha_n!}$$
where the sum runs over all $(n+1)$-tuples of integers 
$\alpha =(\alpha_0 ,...,\alpha_n)$ with $\alpha_0+\dots +\alpha_n=k$ and 
$\alpha_1+2\alpha_2+\dots +n\alpha_n=n$. Notice that we distinguish carefully 
between ${\mathbf S}_n^k$ and ${S}_n^k$, the latter standing for the $k$-th 
power of $S_n$.
The following examples will be useful in the sequel: 
${\mathbf S}_n^0=\delta_{n,0}1$, ${\mathbf S}_0^k=Q_0^k/k!$,
${\mathbf S}_n^1=Q_n$, ${\mathbf S}_1^k=Q_0^{k-1}Q_1/(k-1)!$,
${\mathbf S}_2^2=Q_0 Q_2 + Q_1^2/2$.

Generating series are a useful tool to handle computations with the 
${\mathbf Q}_n^k$ and the ${\mathbf S}_n^k$. Consider for example the series 
${\mathbf S}_\bullet(a+b):=e^{(a+b){\mathbf Q}_\bullet}$: its expansion 
as a series in the variables $a$ and $b$ yields:
\begin{prop} We have, for all $k,l\ge0$:
\label{propLkn}
\begin{eqnarray*}
\binom{k+l}{k} {\mathbf S}_n^{k+l} &=& \sum_{m=0}^n {\mathbf S}_m^{k} 
{\mathbf S}_{n-m}^{l}.
\end{eqnarray*}
In particular
\begin{eqnarray*}
{\mathbf S}_n^{k} &=& \frac{1}{k} \sum_{m=0}^n Q_m {\mathbf S}_{n-m}^{k-1}.
\end{eqnarray*}
\end{prop}

The Hopf algebraic properties of symmetric functions were 
recently exploited with great profit by 
Fauser and coll.~\cite{FauserJarvis04,FauserJarvis06}.
Similarly, we put a simple Hopf algebraic structure on $q\SYM$
uniquely defined by requiring 
the power sums symmetric functions (i.e. $Q_n$ for $n>0$) to form, 
together with $Q_0$, a series of primitive elements or, equivalently, 
by requiring the series 
${\mathbf S_\bullet}$ to be a group-like element. In other terms, the coproduct $\Delta$ on $q\SYM$ is fully specified by requiring that $\Delta (Q_n):=Q_n\otimes 1+1\otimes Q_n,\ n\geq 0$. In particular, the 
coproduct is compatible with the two graduations. When expliciting this 
property in ${\mathbf S}_n^k$, we get:

\begin{prop}
\label{propDeltaLkn}
The coproduct of ${\mathbf S}_n^{k}$ is
\begin{eqnarray*}
\Delta {\mathbf S}_n^{k} &=& \sum_{m=0}^n \sum_{i=0}^k
  {\mathbf S}_m^{i}\otimes {\mathbf S}_{n-m}^{k-i},
\end{eqnarray*}
and its iterated coproduct is
\begin{eqnarray*}
\Delta^{p-1} {\mathbf S}_n^{k} &=& 
\sum_{\substack{n_1+\dots+n_p=n \\ k_1+\dots+k_p=k}} 
  {\mathbf S}_{n_1}^{k_1}\otimes \dots\otimes {\mathbf S}_{n_p}^{k_p},
\end{eqnarray*}
\end{prop}

Note that propositions \ref{propLkn} and \ref{propDeltaLkn}
still hold if the variables $Q_n$ do not commute.

\subsection{On Hopf algebra derivations}

Let $H=\bigoplus\limits_{n\in\N}H_n$ be an arbitrary connected graded 
commutative Hopf algebra and $\deri_0,\deri_1,...,\deri_n,...$ an 
arbitrary sequence of derivations on $H$ with degrees
$0,1,...,n,...$.
That is, for any $p$, the restriction of $\deri_n$ to $H_p$ is a linear map from 
$H_p$ to $H_{p+n}$, and $\deri_n$ satisfies the Leibniz rule: for any $h,l$ in 
$H$, $\deri_n(h\cdot l)=\deri_n(h)\cdot l+h\cdot \deri_n(l)$.
We also assume that the $\deri_n$ commute, so that
the $\deri_n$ generate a commutative subalgebra $\mathcal D$ of $End(H)$ 
(for the composition of maps). Of course, we have in mind the particular derivations $A_m$ acting on $S(U)$, but the following results hold in full generality.

There is therefore, since $q\SYM$ is free 
over the $Q_n$, a universal algebra map $\beta$ from $q\SYM$ to $End(H)$ 
obtained by mapping $Q_n$ to $\deri_n$. We write ${\mathbf L}_\bullet$ for the 
image of ${\mathbf S}_\bullet$ under this map,
and $L_n^k$ for the image of $\mathbf{S}_n^k$.
Note that, for any $p$, $L_n^k$ maps $H_p$ to $H_{p+n}$.
Of course, the identities that hold in $q\SYM$ for the variables $Q_n$ and 
$\mathbf{S}_n^k$ also hold in $End(H)$ for  the variables
$\deri_n$ and $L_n^k$. More surprisingly however, the coalgebra structure 
of $q\SYM$ reflects the action of $\mathcal D$ on $H$.
We refer to \cite{PR2002,EGP1} for similar phenomena occurring in the study 
of Lie idempotents and renormalization in perturbative QFT.

\begin{prop}
We have, for any $X\in q\SYM$ and any $h,h'\in H$:
$$\beta (X)(hh')=\beta (X_{(1)})(h)\beta (X_{(2)})(h')$$
where $X_{(1)}\otimes X_{(2)}$ stands, as usual, for the coproduct of $X$ in 
$q\SYM$.
\end{prop}

The identity can be generalized by a straightforward recursion to compute 
$\beta (X)(h_1...h_n)$.
Notice first that the identity in the Proposition is obvious when $X$ is 
a $Q_n$, since $\beta (Q_n)=\deri_n$ is, by hypothesis, a derivation. Now, 
assume that for $X$ and $Y$ in $q\SYM$ and arbitrary $h,h',l,l'\in H$ 
the above formula holds, that is:
$$\beta (X)(hh')=\beta (X_{(1)})(h)\beta (X_{(2)})(h'),\ \beta (Y)(ll')=
\beta(Y_{(1)})(l)\beta (Y_{(2)})(l').$$ It follows that:
\begin{eqnarray*}
\beta(Y)\circ \beta(X) (hh') &=& \beta(Y) (\beta (X_{(1)})(h)
\beta (X_{(2)})(h'))\\
&=&\beta(Y_{(1)})\circ\beta (X_{(1)})(h)\,
\beta (Y_{(2)})\circ\beta (X_{(2)})(h')\\
&=&\beta((YX)_{(1)})(h)\beta((YX)_{(2)})(h').
\end{eqnarray*}
In other terms, if two elements in $q\SYM$ satisfy the identity in the 
Proposition, their product also satisfies the identity.
Since the $Q_n$ satisfy the identity, and since their products span 
$q\SYM$, the Proposition follows.

Let us consider the particular example of $S(U)$ and of the derivations
$A_{m-1}$. Let us write $\deri_{m-1}(u)=u_{m-1,1}\vee\dots \vee u_{m-1,m}$, 
with an enhanced Sweedler-type notation for the action of $A_{m-1}$ from 
$U$ to $S^m(U)$. 
We get:
\begin{eqnarray*}
L_n^{k}(u) &=& \frac{1}{k} \sum_{m=1}^{n+1} L_{n-m+1}^{k-1} \deri_{m-1}(u)\\
&=&  \frac{1}{k} \sum_{m=1}^{n+1} 
\sum_{\substack{k_1+\dots+k_m=k-1\\n_1+\dots+n_m=n-m+1}}
  L_{n_1}^{k_1}(u_{m-1,1})\vee \dots\vee L_{n_m}^{k_m}(u_{m-1,m}),
\end{eqnarray*}
where $\vee$ is the symmetric product in $S(U)$.
This is a generalization of lemma 13 in \cite{Mestre}
and of proposition 15 in \cite{Mestre2} where Mestre and Oeckl
studied the case where $\deri_n=0$ for $n\not= 0,1$. 
Note that our notation is different from theirs.

%%%%%%%%%%%%%%%%%%%%%%%%%%%%%%%%%%%%%%%%%%%%%%%%%%%%%%%%%%%%%%%%%%%%%%%%%%%
%New section
%%%%%%%%%%%%%%%%%%%%%%%%%%%%%%%%%%%%%%%%%%%%%%%%%%%%%%%%%%%%%%%%%%%%%%%%%%%

\section{One-particle irreducible decompositions}
We consider now the derivation of an explicit decomposition
of a connected Green function into 1PI Green functions
for a general state.

As we have just noticed, the map $Q_m\mapsto \deri_m$ enables us to define $L_n^k$ acting on $S(U)=S(S(V))$. For example
$L_0^{0}(u) = u$, $L_m^{1}(u) = \deri_m (u)$ and
\begin{eqnarray*}
L_0^{2}(u) &=& \frac{1}{2} \sum\limits_{x_1,x_2} D_1(x_1) D_1(x_2)
  \frac{\overline\partial^2 u}{\partial\varphi(x_1)\partial\varphi(x_2)},\\
L_1^{2}(u) &=& \sum\limits_{x_1,x_2,x_3} D_2(x_1,x_2) D_1(x_3)
  \frac{\overline\partial u\is1}{\partial\varphi(x_1)} \vee
  \frac{\overline\partial^2 u\is2}{\partial\varphi(x_2)\partial\varphi(x_3)},\\
L_2^{2}(u) &=& \frac{1}{2}
  \sum\limits_{x_1,x_2,x_3,x_4} D_2(x_1,x_2)D_2(x_3,x_4) 
  \frac{\overline\partial u\is1}{\partial\varphi(x_1)}\vee
  \frac{\overline\partial^2 u\is2}{\partial\varphi(x_2)\partial\varphi(x_3)}
  \vee \frac{\overline\partial u\is3}{\partial\varphi(x_4)}
  \\ &&
  + \frac{1}{2}\sum\limits_{x_1,x_2,x_3,x_4} D_3(x_1,x_2,x_3)D_1(x_4)
  \frac{\overline\partial u\is1}{\partial\varphi(x_1)}\vee
  \frac{\overline\partial u\is2}{\partial\varphi(x_2)}\vee
  \frac{\overline\partial^2 u\is3}{\partial\varphi(x_3)\partial\varphi(x_4)},\\
L_1^{3}(u) &=& \frac{1}{2}
  \sum\limits_{x_1,x_2,x_3,x_4} D_1(x_1)D_1(x_2) D_2(x_3,x_4)
  \frac{\overline\partial^2 u\is1}{\partial\varphi(x_1)\partial\varphi(x_3)}
  \vee
  \frac{\overline\partial^2 u\is2}{\partial\varphi(x_2)\partial\varphi(x_4)}
  \\&&
  + \frac{1}{2}
  \sum\limits_{x_1,x_2,x_3,x_4}  D_1(x_1) D_1(x_2) D_2(x_3,x_4)
  \frac{\overline\partial^3 u\is1}{\partial\varphi(x_1)\partial\varphi(x_2)
  \partial\varphi(x_3)} \vee
  \frac{\overline\partial u\is2}{\partial\varphi(x_4)}.
\end{eqnarray*}

\subsection{A tree interpretation}
The operator $L_n^{k}$ can be written as a sum over all the bipartite trees
with $k$ white vertices and $n+1$ black vertices.
This description in terms of trees is important because, in standard QFT,
a connected Green function can also be described as a tree of
1PI Green functions --a description we want to extend to the case of initial correlations. 
To give a more precise relation between
$L_n^{k}$ and bipartite trees, we consider the expression
for $L_n^{k}$ in terms
of partitions $\alpha$:
$$L_n^k=\sum_\alpha\frac{\deri_0^{\alpha_0}}{\alpha_0 !}\dots 
\frac{\deri_n^{\alpha_n}}{\alpha_n !},$$
where the sum runs over the sequences $\alpha$ of 
nonnegative integers with 
$\alpha_0+\dots +\alpha_n=k$ and $\alpha_1+\dots +n \alpha_n=n$.
The monomial corresponding to a given $\alpha$
is represented by the sum of all bipartite trees
with $k$ white vertices and $n+1$ black vertices,
such that $\alpha_i$ white vertices have valency $i+1$,
for $i=1,\dots, n$. 

The terms of lowest degrees are 
 \begin{eqnarray*}
L_0^0 &=& \id = \tunn,\\
L_0^1 &=& \deri_0 = \tdeux\, ,\\
L_0^2 &=& \frac{1}{2!} \deri_0^2 = \ttroisb\, ,\\
L_1^1 &=& \deri_1 = \ttroisn\, ,\\
L_0^3 &=& \frac{1}{3!} \deri_0^3 = \tquatreb\, ,\\
L_1^2 &=& \deri_0 \deri_1 = \tquatrel\, ,\\
L_2^1  &=& \deri_2 = \tquatren\, ,\\
L_0^4 &=& \frac{1}{4!} \deri_0^4 = \tcinqb\, ,\\
L_2^2 &=& \deri_0 \deri_2 + \frac{1}{2} \deri_1^2 = \tcinqfn + \tcinqln\, ,\\
L_1^3 &=& \frac{1}{2} \deri_0^2 \deri_1 = \tcinqlb + \tcinqfb\, ,\\
L_3^1  &=& \deri_3 = \tcinqn\, ,\\
L_0^5 &=& \frac{1}{5!} \deri_0^5 = \tsixb\, ,\\
L_3^2 &=& \deri_0 \deri_3 + \deri_1 \deri_2 = \tsixftn+\tsixfn\, ,\\
L_2^3 &=& \frac{1}{2!} \deri_0^2 \deri_2 + 
   \frac{1}{2!} \deri_0\deri_1^2 = \tsixgfb+\tsixff + \tsixl+\tsixgfn \, ,\\
L_1^4 &=& \frac{1}{3!} \deri_0^3 \deri_1 = \tsixftb+\tsixfb\, ,\\
L_4^1  &=& \deri_4 = \tsixn\, .
\end{eqnarray*}

To calculate the value of a tree of $L^k_n$:
(i) Associate to each of the $k+n$ edges a variable
$x_i$, with $i=1,\dots,k+n$.
(ii) To each white vertex $v$,
  associate the factor $D_m(x_{i_1},\dots,x_{i_m})$,
  where $m$ is the valency of $v$ and
  $x_{i_1},\dots,x_{i_m}$ are the variables associated
  to the edges incident to $v$.
(iii) There are $n+1$ black vertices. Split $u$ into
  $n+1$ parts by
   $\Deltas^{n}u = u\is1\otimes\dots\otimes u\is{n+1}$.
  Number the black vertices from 1 to $n+1$ and to vertex
  $\ell$ associate the factor
\begin{eqnarray*}
  \frac{\overline\partial^m u\iis{\ell}}{\partial \varphi(x_{i_1})\dots
  \partial\varphi(x_{i_\ell})},
\end{eqnarray*}
where $x_{i_1},\dots,x_{i_\ell}$ are the variables associated
  to the edges incident to the black vertex number $\ell$.
(iv) Multiply the factors corresponding to the black
vertices with the product $\vee$ in $S(U)$.
(v) Divide the
resulting value by the order of the symmetry
group of the tree.
 
\subsection{The 1PI components of forms}

The last step before we can write a connected form in terms
of 1PI forms is to give a reasonable definition of
what is the 1PI component of a form, similarly to the definition of the 
connected components of forms. Several definitions are possible.
The simplest one was proposed by Hall \cite{Hall}
and has recently provided detailed structural results \cite{BrouderBSL}.
Here we consider a definition which is strictly more general
than Hall's and that leads to an interesting structure.
In a graph, it is easy to describe what we mean by cutting a line
or a set of lines; this approach leads, in classical QFT 
(with 2-point Feynman propagators) to the definition of 1PI Feynman 
diagrams as connected diagrams that are still connected when an arbitrary 
propagator line is cut. We propose to generalize the notion by replacing 
the Feynman form (that is, the classical case where only 2-point
Feynman propagators are considered) by an arbitrary unital
form. 

Our approach is rooted in the Hopf algebraic picture of QFT. Notice 
however that our constructions could be translated \it mutatis mutandis 
\rm in the language of functional derivatives. For example, the 
derivatives $\frac\partial{\partial \varphi (x)}$ that we have used 
in the definition of the operators $\deri_i$ were defined as usual derivatives 
(in the polynomial algebra over the symbols $\varphi (x)$) but could be 
understood alternatively as functional derivatives. The same observation 
holds for our forthcoming constructions. 

In proposition~\ref{proptuples}, for any
$u=\varphi^{n_1}(x_1)...\varphi^{n_k}(x_k)$, we have expanded $\rho(u)$ 
as a linear combination of terms such as
$r(\varphi^{n_1^{1}}(x_1)
\dots \varphi^{n_k^{1}}(x_k))...r(\varphi^{n_1^{l}}(x_1)\dots 
\varphi^{n_k^{l}}(x_k)).$
Let us consider a connected term $$t:=r(\varphi^{n_1^{1}}(x_1)
\dots \varphi^{n_k^{1}}(x_k))...r(\varphi^{n_1^{l}}(x_1)\dots 
\varphi^{n_k^{l}}(x_k))$$ in 
$\ee^{*r}(u)$.

\begin{de}
The connected term $t$ is said to be one-particle reducible if and only if, 
there exists $i \in \{1,...,l\}$ and $\{i_1,...,i_p\}\subset \{1,...,k\}$ 
such that 
\begin{enumerate}
\item $\varphi^{n_1^{i}}(x_1)
\dots \varphi^{n_k^{i}}(x_k)=\varphi(x_{i_1})
\dots \varphi(x_{i_p})$ 
\item furthermore, in the remaining part 
of $t$, 
\begin{multline*}
r(\varphi^{n_1^{1}}(x_1)
\dots \varphi^{n_k^{1}}(x_k))...r(\varphi^{n_1^{i-1}}(x_1)\dots 
\varphi^{n_k^{i-1}}(x_k))\\r(\varphi^{n_1^{i+1}}(x_1)\dots 
\varphi^{n_k^{i+1}}(x_k))...
r(\varphi^{n_1^{l}}(x_1)\dots 
\varphi^{n_k^{l}}(x_k))
\end{multline*} 
the connected components of 
$x_{i_1},...,x_{i_p}$ are either empty or pairwise disjoint.
\end{enumerate}
A connected term that is not one-particle reducible is said to be one-particle irreducible (1PI).
\end{de}

For example, $r(\varphi(x_1)\varphi(x_2))^2$ (a loop constructed out of two 
two-point propagators) is 1PI (in our situation, and also in the usual 
picture), and so is $$r(\varphi(x_1)\varphi(x_2))r(\varphi(x_1)\varphi(x_3))
r(\varphi(x_1)\varphi(x_2)\varphi(x_3)),$$ whereas
$$r(\varphi(x_1)\varphi(x_2))r(\varphi(x_1)\varphi(x_3))
r(\varphi(x_1)\varphi(x_2)\varphi(x_4))$$
or $$r(\varphi(x_1)\varphi(x_2))r(\varphi(x_1)\varphi^2(x_3))
r(\varphi(x_2)\varphi^2(x_4))$$ are not.

\begin{de}
The 1PI component $\rhoi$ of a form
is the sum of all the 1PI terms 
in the connected component of $\rho$.
\end{de}

Recall that forms, their connected components, and also their 1PI components 
are linear maps from $U=S(V)$ to $\C$. However, any such linear map $l$ 
can be uniquely extended to a multiplicative map, still written $l$ from 
$S(U)$ to $\C$ (that is, to a character of the algebra $S(U)$, in the 
algebraic terminology). Concretely, for $u_1,...,u_n\in U, l(u_1\vee 
\dots\vee u_n):=l(u_1)...l(u_n)$. In particular, the connected and 1PI
% and 1PI to all finite orders 
components of forms can be viewed as characters 
of the algebra $S(U)$, so that, for example, an expression such as 
$\rho_c\circ \deri_m$ makes sense as the composition of a derivation of 
$S(U)$ and a map from $S(U)$ to $\C$.

Let us consider a simple example to illustrate these ideas, namely the 
structure equation linking connected and 1PI components of forms in 
the most common picture of pQFT: an interacting theory --say $\varphi^3$--
with Feynman diagrams built of 3-valent interaction vertices and 2-point
propagators. A general Feynman diagram can be described as 1PI diagrams 
connected by $n\in \N$ Feynman propagators satisfying the property that 
cutting any of these propagators makes the original diagram disconnected. 
Taking into account the symmetry factor $n!$ arising from the fact that 
these Feynman propagators can be cut successively in an arbitrary order 
results into a structure equation relating the connected and 
1PI Green functions. In Hopf algebraic terms:
$$\rho_c=\rhoi\circ e^F$$
where $F= \deri_{1}$ is the derivation of $S(U)$ associated to the Feynman propagator:
for all $ u_1,...,u_n\in U=S(V)$, 
\begin{eqnarray*}
F(u_1\vee\dots \vee u_n) &=& \frac{1}{2}
\sum\limits_{x\not= y}\sum\limits_{i\leq n}  D_2(x,y) u_1\vee\dots \vee 
u_{i-1}\vee \\
&&(\frac{\overline\partial u_{i,\{1\}}}{\partial\varphi (x)}\vee 
\frac{\overline\partial u_{i,\{2\}}}{\partial\varphi (y)})\vee u_{i+1}\vee \dots 
\vee u_n,
\end{eqnarray*}
with $D_2$ the (quasi-)free 2-point Green function. 

In the general case, replacing Feynman propagators $D_2$
by arbitrary propagators $D_n$, doesn't change the general 
principles of the proof. An arbitrary Feynman diagram for an interacting 
theory as the ones considered previously in the present section can still be 
cut into 1PI pieces connected by a family of $n$-point propagators, $n\in N$, 
in such a way that removing any of these $n$-point propagators splits the 
original diagram into $n$ connected pieces. For a given $n$, the associated 
symmetry factor is $p_n!$, where $p_n$ is the number of $n$-point propagators 
in the family. These observations result in a family of structure
identities for 1PI diagrams at all orders, and an identity that 
should be understood as a structure theorem for the perturbative approach 
to interacting theories.

\begin{thm} (Structure of connected forms)
\label{Master}
For an arbitrary connected form $\rhoc$, we have
$$\rho_c=\rhoi\circ {\bf L}_\bullet =\rhoi\circ 
e^{\left( \sum\limits_{n\in \N}\deri_n\right)},$$
and
$$\rhoi=\rho_c\circ e^{\left( -\sum\limits_{n\in \N}\deri_n\right)}.$$
\end{thm} 

We remark that the effect of $\deri_0$ is just a shift of the fields:
for instance $\ee^{\deri_0}(\varphi^n(x)) = (\varphi(x)+D_1(x))^n$.
For $u=\varphi^{n_1}(x_1)\dots\varphi^{n_p}(x_p)$, we have
$\deri_m(u)=0$ if $m\ge p$ because $\deri_m$ splits $u$ into $m+1$ pieces
and the coordinates $x_i$ of these pieces must all be different.
More generally, $L_n^k(u)$ vanishes if $n\ge p$.
Because of the trivial effect of $\deri_0$ we put $\deri_0=0$ and we get 
\begin{eqnarray*}
\rhoc(u) &=& \rhoi(u) + \sum_{n=1}^{p-1}\sum_{k=1}^{n_1+\dots+n_p} 
   \rhoi(L_n^k(u)).
\end{eqnarray*}
In particular, 
$\rhoc(\varphi^n(x))=\rhoi(\varphi^n(x))$ and, 
for $x\not=y$,
\begin{eqnarray*}
\rhoc(\varphi^n(x)\varphi^m(y)) &=&
\rhoi(\varphi^n(x)\varphi^m(y))
+mn D_2(x,y)
\rhoi(\varphi^{n-1}(x))
\rhoi(\varphi^{m-1}(y)).
\end{eqnarray*}
The first equation of theorem \ref{Master} describes the
connected Green functions in terms of 1PI Green functions.
It is an extension  to general states of the standard
QFT result and of a theorem by Mestre and Oeckl \cite{Mestre2}.
The second equation is new even in the QFT context: it describes
the 1PI Green functions as a linear combination of
products of connected Green functions. In QFT, 1PI Green functions
are expressed in terms of amputated connected Green functions.
Here, we do not amputate the Green functions (this is not
allowed for a general state because parts of the Green
functions belong to the kernel of the differential operator
used in the equation of motion of the free field).

The consequences of these identities for the QFT of interacting systems, 
and the fine study of connected and 1PI amplitudes are postponed to 
further work.

\section{Conclusion}
In this paper, we developed mathematical tools to extend the
relation between connected Green functions and 1PI Green functions
from the case of a quasi-free ground state
to the case of a general state.  Our main result is the
structure of connected forms described by theorem \ref{Master}. 
This work can be extended
in two directions. On the physical side, the main structure identity can be used
to derive resummation theorems that generalize Friedberg's formulas
\cite{Friedberg}. On the mathematical side, many of our results
can be extended to the case of non commuting variables.

\section{Acknowledgements}
This work originated as a part of a long term joint project on the
combinatorics of QFT with Alessandra Frabetti. 
We thank her very warmly: without the many
intense and inspiring discussions that we had with her, 
this paper would not exist.

We thank Raymond Stora for numerous fruitful comments on a previous
version of the manuscript. We also thank warmly Achim Randelhoff
who pointed out to us that the definition of the derivations $\deri_{m-1}$ in a previous version of this article was not suited to their later application to 1-particle irreducibility.

We are grateful to Muriel Livernet for an inspiring remark
about the relation between Ref.~\cite{Mestre} and one of her
works \cite{Livernet} and a proof that some results of
Ref.~\cite{Mestre} are valid in a general algebraic setting.
The help of Jean-Louis Loday is gratefully acknowledged 
by Ch. B., who is also grateful to John Challifour for sending a
copy of his unpublished paper with Wightman.

%\bibliographystyle{unsrt}
%\bibliography{qed}

\end{document}